\documentclass[structabstract]{aa}

\usepackage{txfonts}
\usepackage{natbib}
\usepackage{graphicx}
\usepackage{color}
\usepackage{longtable,lscape}

\bibpunct{(}{)}{;}{a}{}{,} % to follow the A&A style

\newcommand{\expo}[1]{$10^{#1}$}
\newcommand{\texpo}[1]{$\times 10^{#1}$}

\newcommand{\kmpers}{$\mathrm{km \, s^{-1}}$}

\newcommand{\adegdot}[2]{\mbox{#1$\stackrel {\circ}{_{\bf \cdot}}$#2}}
\newcommand{\amindot}[2]{\mbox{#1$\stackrel {\prime}{_{\bf \cdot}}$#2}}

\newcommand{\orthowater}{\textit{ortho}-water}     %molecules
\newcommand{\ohtvao}{\textit{o}-H$_2$O}

\newcommand{\phtva}{\textit{p}-H$_2$} 
\newcommand{\htvao}{H$_2$O}  
\newcommand{\htva}{H$_2$}  

\newcommand{\amin}{$^{\prime}$}                   %arcus and coordinates
\newcommand{\asec}{$^{\prime \prime}$}
\newcommand{\adeg}{$^{\circ}$}

\newcommand{\atwozero}{$\alpha_{2000}$}
\newcommand{\dtwozero}{$\delta_{2000}$}

\newcommand{\msun}{$M_{\odot}$}

\newcommand{\mdot}{{\it \.{M}}}
\newcommand{\msunyr}{$M_{\odot} \, {\rm yr}^{-1}$}

\newcommand{\radex}{\texttt{RADEX} }            % programs

\newcommand{\swas}{SWAS}                        %observatories
\newcommand{\odin}{Odin}
\newcommand{\sest}{SEST}
\newcommand{\iso}{ISO}

\begin{document}

\title{\odin\ observations of water in molecular outflows and
  shocks\thanks{\odin\ is a Swedish-led satellite project funded jointly
    by the Swedish National Space Board (SNSB), the Canadian Space
    Agency (CSA), the National Technology Agency of Finland (Tekes)
    and Centre National d’Etude Spatiale (CNES).}$^{\rm ,}$\thanks{The
    Swedish ESO Submillimetre Telescope (SEST) located at La Silla,
    Chile was funded by the Swedish Research Council (VR) and the
    European Southern Observatory. It was decommissioned in 2003.}}

\author{P. Bjerkeli\inst{1}
     \and R. Liseau\inst{1}
     \and M. Olberg\inst{1,2}
    \and E. Falgarone \inst{3}
    \and U. Frisk \inst{4}
    \and \AA. Hjalmarson\inst{1}
    \and A. Klotz \inst{5}
    \and B. Larsson \inst{6}
    \and A.O.H. Olofsson \inst{7,1}
    \and G. Olofsson \inst{6}
    \and I. Ristorcelli \inst{8}
    \and Aa. Sandqvist \inst{6}}

\offprints{\\ P. Bjerkeli, \email{per.bjerkeli@chalmers.se}}

\institute{Onsala Space Observatory, Chalmers University of Technology, SE-439 92 Onsala, Sweden
\and SRON, Landleven 12, P.O.Box 800, NL-9700 AV Groningen, The Netherlands
\and Laboratoire de Radioastronomie - LERMA, Ecole Normale Superieure, 24 rue Lhomond, 75231 Paris Cedex 05, France
\and Swedish Space Corporation, PO Box 4207, SE-171 04 Solna, Sweden
\and CESR, Observatoire Midi-Pyr\'en\'ees (CNRS-UPS), Universitet\'e de Toulouse, BP 4346, 31028 Toulouse Cedex 04, France
\and Stockholm Observatory, Stockholm University, AlbaNova University Center, SE-106 91 Stockholm, Sweden
\and GEPI, Observatoire de Paris, CNRS, 5 Place Jules Janssen, 92195 Meudon, France
\and CESR, 9 avenue du Colonel Roche, BP 4346, 31029 Toulouse, France}

%\date{draft February, 2009}

\abstract {} {We investigate the \orthowater\ abundance in outflows
  and shocks in order to improve our knowledge of shock chemistry and
  of the physics behind molecular outflows.} {We have used the \odin\
  space observatory to observe the \htvao($1_{10}-1_{01}$) line. We
  obtain strip maps and single pointings of 13 outflows and two
  supernova remnants where we report detections for eight sources. We
  have used \radex\ to compute the beam averaged abundances of
  \ohtvao\ relative to \htva. In the case of non-detection, we derive
  upper limits on the abundance.}{Observations of CO emission from the
  literature show that the volume density of \htva\ can vary to a
  large extent, a parameter that puts severe uncertainties on the
  derived abundances. Our analysis shows a wide range of abundances
  reflecting the degree to which shock chemistry is affecting the
  formation and destruction of water. We also compare our results with
  recent results from the \swas\ team.}{Elevated abundances of
  \orthowater\ are found in several sources. The abundance reaches
  values as high as what would be expected from a theoretical C-type
  shock where all oxygen, not in the form of CO, is converted to
  water. However, the high abundances we derive could also be due to
  the low densities (derived from CO observations) that we assume.
  The water emission may in reality stem from high density regions
  much smaller than the \odin\ beam. We do not find any relationship
  between the abundance and the mass loss rate. On the other hand,
  there is a relation between the derived water abundance and the
  observed maximum outflow velocity.}

\maketitle 

\section{Introduction}
Deeply embedded Class 0 stellar systems are observed to be associated
with high velocity bipolar outflows \citep[see e.g.][]{Snell:1980lr}
which are believed to play an important role when stars are
formed. During the phase when material is accreted onto the newborn
star through the circumstellar disk, outflows are responsible for a
necessary re-distribution of angular momentum. The specific angular
momentum of the infalling material must at some point
decrease to allow the final collapse. Although the basic theoretical
concepts can be understood, there is still a great observational need
to get further knowledge about the engine of these flows. Different
models describing the driving mechanisms have been proposed and for
that reason it is important to derive abundances of different species
in order to distinguish between different physical scenarios. In this
context, water is interesting in the sense that it is strongly
affected by the presence of different types of shocks. At low
temperatures, water is formed through a series of ion molecule
reactions. This process is relatively slow and enhanced water
abundances are thus not expected. At higher temperatures, the
activation barrier for neutral-neutral reactions is reached, and for
that reason water can be formed in a much more efficient way. Such
elevated temperatures are reached in low velocity shocks, where the
shock is smoothed by friction between ions and neutrals, called
continuous shocks \citep[see e.g.][]{Bergin:1998lr}. Here, \htva\ is
prevented from destruction and enhanced water abundances are
expected. In this scenario, water is not only formed through reactions
with oxygen but can also be released from its frozen state on dust
grains \citep{Kaufman:1996qy}. In the discontinuous type of shock
(jump shock), \htva\ is instead dissociated and water formation is
prevented. The detection of water is aggravated by the difficulty
of observing from ground based observatories. Prior to the
launch of \odin\ \citep{Nordh:2003qy,Hjalmarson:2003kx}, two space
born observatories capable of detecting water have been in operation,
the \textit{Infrared Space Observatory} (\iso) \citep{Kessler:1996fj}
and the \textit{Submillimeter Wave Astronomy Satellite} (\swas)
\citep{Melnick:2000fk}. The latter of these two also had the ability
to observe the ground state transition of \ohtvao\, although the beam
size was larger (3\amindot\, 3 $\times$ 4\amindot\, 5 elliptical
compared to 2\amindot\, 1 circular for Odin). Water abundances derived
from \iso\ data are in general higher than those derived from \swas\
data. This discrepancy is addressed in the paper by
\citet{Benedettini:2002fk}. In 2003, the \textit{Spitzer Space
  Telescope}, capable of detecting hot water was launched
\citep{Werner:2004fk}.

\begin{table*}
\begin{flushleft}
\caption{Observation log for the sources analyzed in this paper.}
\label{tab:table1}
  \renewcommand{\footnoterule}{}  % to avoid a line before footnotes
  \begin{tabular}{l c c r c c r l l}
\hline
\hline
\noalign{\smallskip}
    Source & $\alpha$(2000) & $\delta$ (2000) & Distance & Backend & Date & $t_{\textrm{\tiny int}}$ &\\
    & (hr:min:sec) & (deg:min:sec) & (pc) & & (YYMMDD) & (hr) &\\% \bf{(molecule/continuum)} \\
\noalign{\smallskip}
    \hline
\noalign{\smallskip}
\noalign{\smallskip}
L1448   &  03:25:38.8 & +30:44:05 &250 & AOS & 060722 - 060813  & 38 &\\
HH 211  & 03:43:56.5 & +32:00:50 &315 & AOS & 070119 - 070209 & 22  &\\
L1551& 04:31:34.1 & +18:08:05 & 140 & AC2 & 020208 - 020213  & 12 &\\
IC443-G & 06:16:43.4 & +22:32:24 & 1500 & AOS & 020319 - 020321  & 5 &\\    
TW Hya  & 11:01:51.9 & $-$34:42:17 & 56 & AOS & 031118 - 031204  & 18 &\\
$\epsilon$ Cha\,{\sc I\,}N &11:09:54.6 & $-$76:34 24  &150 & AC2 & 031205 - 031212 & 11 &\\
Sa136 (BHR71)  & 12:01:37.0 & $-$65:08:54 & 200 &AOS & 060528 - 060721 & 41 &\\
HH54 B  & 12:55:50.3 & $-$76:56:23 &200 & AC2 & 050502 - 050620 & 10 &\\
G327.3-0.6& 15:53:08.7 & $-$54:37:01& 2900&AC2&030731 - 030807&11 &\\
NGC6334\,{\sc I}& 17:20:53.4 & $-$35:47:02 & 1700 & AC1 & 010926 - 010927 & 2 &\\
Ser SMM1 & 18:29:49.8 & +01:15:21 & 310 & AC2 &041026 - 041031  & 9 &\\
3C391 BML & 18:49:22.0 & $-$00:57:22 & 8000 & AC2 & 041101 - 041106  & 14 &\\
B335& 19:37:01.0 & +07:34:11 & 250 & AC2 & 031024 - 031101 & 7 &\\
L1157   & 20:39:06.4 & +68:02:13 &250 & AC2/AOS & 051129 - 060127 & 119 &\\
NGC7538 IRS1& 23:13:46.8 & +61:28:10 &2700 & AOS & 011214 - 011214 & 1 &\\
 \noalign{\smallskip}
\noalign{\smallskip}
    \hline
  \end{tabular}

\end{flushleft}

\end{table*}

In this paper, \htvao($1_{10}-1_{01}$) observations of 13 outflows and
two supernova remnants are discussed. Shocks from supernova explosions
have similar effect on the chemical conditions as molecular
outflows. The different sources are discussed in Section
(\ref{subsection:notes}) and summarized in Table~\ref{tab:table1} and
2. Table~3 includes other outflows observed by \odin\ that have
already been investigated by other authors or are in preparation
for publication (W3, Orion KL, $\epsilon$ Cha-MMS1, IRAS
16293-2422, S140 and VLA1623). The analysis carried out in these
papers is however different from the analysis made in the present
paper. Similar observations as the ones discussed here have recently
been presented by the \swas\ team \citep{Franklin:2008fk}. For that
reason we make a brief comparison of the results for common sources.

\section{Observations and reductions}
\subsection{\htvao\ observations}
All \ohtvao\ observations were made with the \odin\ space observatory
between 2002 and 2007 (see Table~\ref{tab:table1}). Each revolution of
96 minutes allows for 61 minutes of observations, whereas the source
is occulted by the Earth for the remaining 35 minutes. The
occultations allow for frequency calibration using atmospheric
spectral lines. At the wavelength of the \orthowater\ ground state
transition, the 1.1\,m Gregorian telescope has a circular beam with
Full Width Half Maximum (FWHM) of 126${}^{\prime \prime}$
\citep{Frisk:2003lr}. The main beam efficiency is close to 90 \% as
measured from Jupiter mappings \citep{Hjalmarson:2003kx}. The main
observing mode was sky switching, where simultaneous reference
measurements from an unfocused 4\adegdot\, 4 FWHM sky beam were
acquired. Position switching, where the entire spacecraft is
re-orientated in order to obtain a reference spectrum, was the method
of observation for a smaller number of targets. Three different
spectrometers were used. Two of these are autocorrelators (AC1, AC2)
and the third one is an acousto-optical spectrometer (AOS). The AOS
has a channel spacing of 620\,kHz (0.33 \kmpers\,at 557 GHz), while
the autocorrelators can be used in different modes. The majority of
the data have a reconstructed pointing offset less than
20\arcsec.
The data processing and calibration is described in detail by
\citet{Olberg:2003fj}.

\section{Results}
The baseline-subtracted \htvao\ spectra for the 15 previously not
published sources are presented in the right column of Figures~
\ref{fig:waterspectra} - \ref{fig:waterspectra4}. All spectra are
smoothed to a resolution of 0.5\,\kmpers. The velocity interval has
been chosen in order to emphasize the line profiles. In the left
column, the calibrated raw data are plotted for comparison. These
spectra have their baselines subtracted using a zeroth order
polynomial.

\section{Discussion}
\label{section:discussion}

\subsection{Densities, temperatures and radiative transfer analysis}
\begin{figure}
  \resizebox{\hsize}{!}{\includegraphics{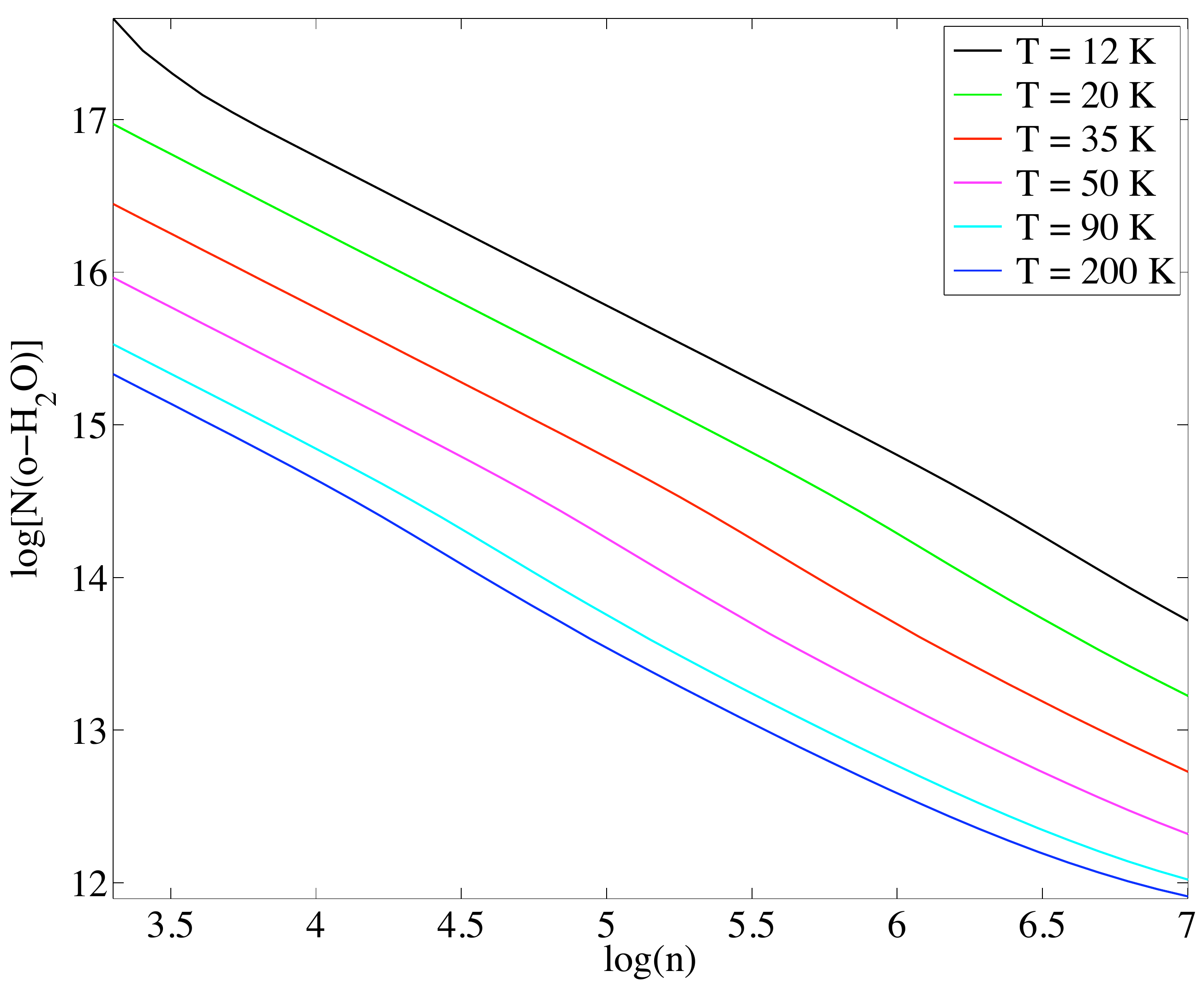}}
  \caption{The derived \ohtvao\ column density as a function of
    volume density for different temperatures. The line intensity for
    this test case has been set to 0.1 K while the line width has been
    set to 10 \kmpers.}
  \label{fig:radex}
\end{figure}

In this paper we derive the beam averaged \orthowater\ abundance. The
beam size is however likely to be larger than the emitting regions for
several of the sources that are analyzed. In order to do this, we use
\radex \footnote{http://www.strw.leidenuniv.nl/~moldata/radex.html}
  \citep{van-der-Tak:2007fk}. It is a publically available code that
  uses the method of mean escape probability for the radiative
  transfer. For interstellar gas at relatively low temperatures and
  densities ($\ll10^{8}$ cm${^{-3}}$) water excitation will be
  subthermal and the emission be on the linear part of the curve of
  growth, i.e. even an optically thick line will behave as being
  effectively optically thin \citep{Linke:1977lr,Snell:2000lr}.  For
  an effectively optically thin line, the result of a radiative
  transfer code like \radex\ is essentially the same as that of the
  analytical expression for a collisionally excited transition
  \begin{equation}
    \label{eq:5}
    F = h \nu \frac{\Omega}{4 \pi} X_{\mathrm{mol}} N(\mathrm{H_2}) \gamma_{\mathrm{lu}} n(\mathrm{H_2}) \frac{\frac{\beta_{\mathrm{e}} n_{\mathrm{c}}}{n(\mathrm{H_2})}}{\frac{\beta_{\mathrm{e}} n_{\mathrm{c}}}{n(\mathrm{H_2})} + \frac{\gamma_{\mathrm{lu}}}{\gamma_{\mathrm{ul}}} + 1}
  \end{equation}
  \citep{Liseau:1999uq}. $F$ is the integrated line
  flux, $\gamma_{\mathrm{lu}}$ and $\gamma_{\mathrm{ul}}$ are the upward
  and downward collision coefficients, $n_{\mathrm{c}}$ is the
  critical density and $\beta_{\mathrm{e}}$ is the photon escape
  probability. The molecular datafiles that are used by \radex\ are
  taken from the Leiden Atomic and Molecular Database
  (LAMDA)\footnote{http://www.strw.leidenuniv.nl/~moldata/}. Here, the
  collision rates between \ohtvao\ and \htva\ have been retrieved from
  \citet{Phillips:1996qy}, \citet{Dubernet:2002lr} and
  \citet{Faure:2007lr}. Recently, new rate coefficients of \ohtvao\
  with \phtva\ have been published by
  \citet{Dubernet:2009lr}. According to the results presented in this
  paper (their Figure~8), collision rates used in \radex\ should not
  differ by more than a factor of three in the temperature regime
  below 40 K. The ortho to para ratio of \htva\ is in \radex\ assumed
  to be thermal. As input parameters, the line intensity, line width,
kinetic temperature and volume density of the observed gas must be
supplied. The difficulty in obtaining good estimates for the two
latter parameters has to be kept in mind. In Figure~\ref{fig:radex},
we plot the \ohtvao\ column density as a function of volume density for a
test case. The line intensity has been set to 0.1 K and the linewidth
to 10 \kmpers. From this figure, it is clear that the derived
abundances are very uncertain when the kinetic temperatures are low.

In this study, the temperature is taken from the literature while the
volume density is inferred from CO observations carried out
by others. These two parameters are however not expected to have
constant values across the large \odin\ beam due to the quite complex
morphology of molecular outflows. For all the sources, we only include
the cosmic microwave background as a radiation field. For simplicity
we have chosen the line intensity to be equal to the peak value while
the line width is taken as the width of the line at 50 \% of this
value. The output column density of \orthowater\ is then used to
derive the abundance relative to molecular hydrogen, $X(o$-$\rm{H_2O})
= \it{N}(o$-$\rm{H_2O})/\it{N}(\rm{H_2})$.

A widely recognized method to obtain the column density for \htva\ is
to measure the CO abundance assuming a constant universal ratio, e.g.
[$\rm{CO/H_2}$] = \expo{-4} \citep[see e.g.][]{Dickman:1978fk}. In
this paper we use this method where it is feasible (different methods
are used for TW Hya and 3C391 BML).
Volume densities and beam averaged column densities are estimated
from literature data assuming cylindrical geometry and a mean
molecular weight $\mu = 2.4$. No correction for inclination is
made. The inferred volume densities (Method 1) should be considered
as lower limits for two reasons. First, the size of the water
emitting regions is poorly known. These regions may very well be
smaller than the CO emitting regions, potentially resulting in a
higher average density. Secondly, shocks, if present, will compress
the gas even further. For some of the sources, there are estimates
of the volume density, given in the literature, that are
significantly higher than the ones used in this paper. In these
cases we also estimate an alternative \orthowater\ abundance (Method
2).

Table~2 includes the measured integrated intensity over the observed
lines and the derived water abundance. The integrated intensity is
measured over the entire line including the central region as well as
the outflow wings. Excepted are those outflow sources, for which
strong self absorption can be seen (e.g. L1157, Ser SMM1). For these
objects, the integrated intensity has been measured for the red and
the blue wings separately. For sources with no detection, we set a $3
\sigma$ upper limit on the integrated intensity in a velocity interval
of 10 \kmpers (except for TW Hya, where a linewidth of 1 \kmpers\ has
been used).

\subsection{Notes on individual sources}
\label{subsection:notes}

\subsubsection{L1448}
\begin{figure}
  \resizebox{\hsize}{!}{\includegraphics{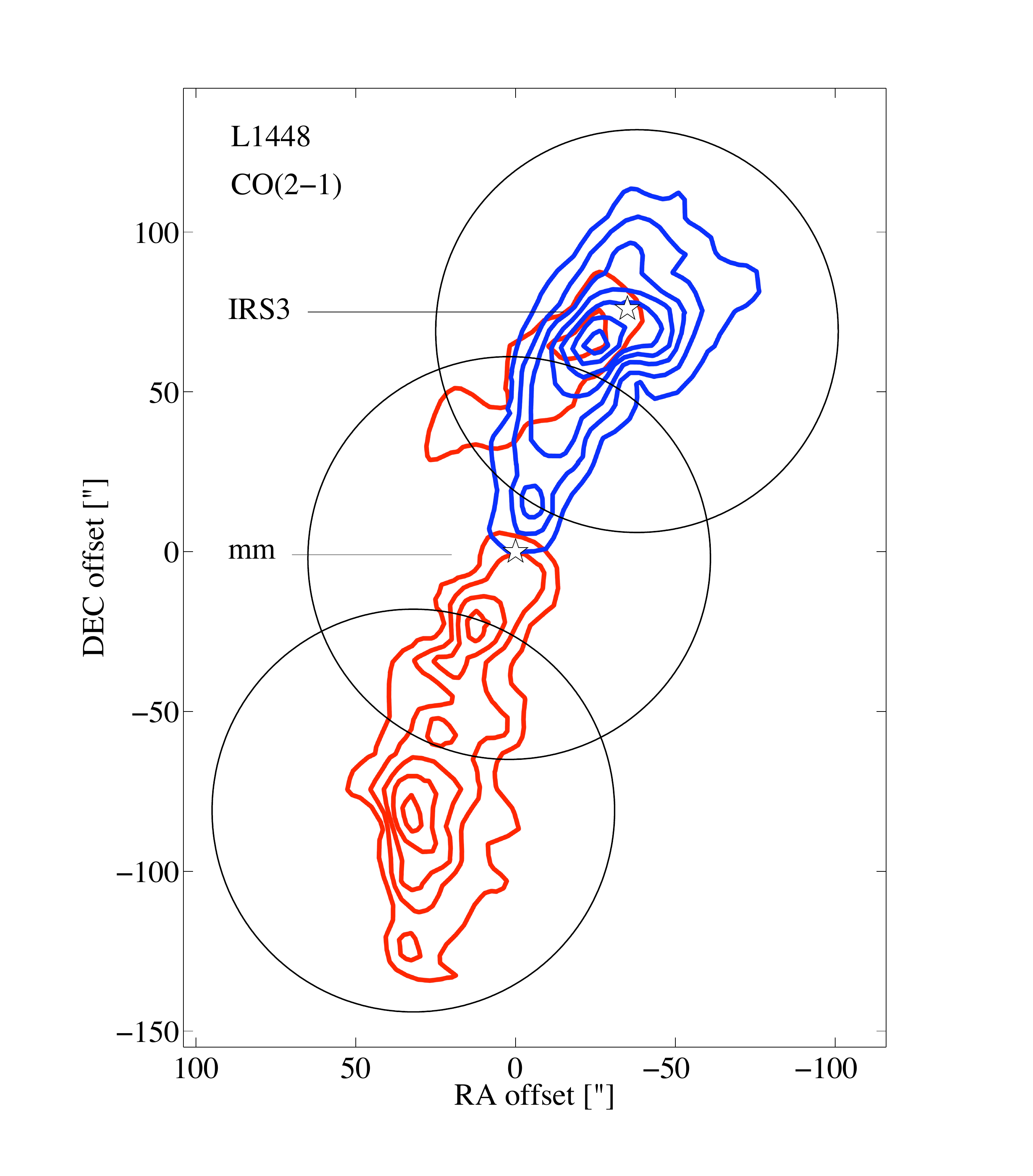}}
  \caption{The three positions observed by \odin\ are shown overlaid
    on a CO (2-1) map of L1448 \citep{Bachiller:1995uq}. The circles
    correspond to the Odin beam at 557\,GHz. Coordinate offsets are
    given with respect to L1448-mm: \atwozero = 03:25:38.8, \dtwozero
    = 30:44:05.0. The positions of L1448-mm and L1448 IRS3 are
    indicated by star symbols.}
  \label{fig:L1448}
\end{figure}
\begin{figure}
\resizebox{0.9\hsize}{!}{\includegraphics{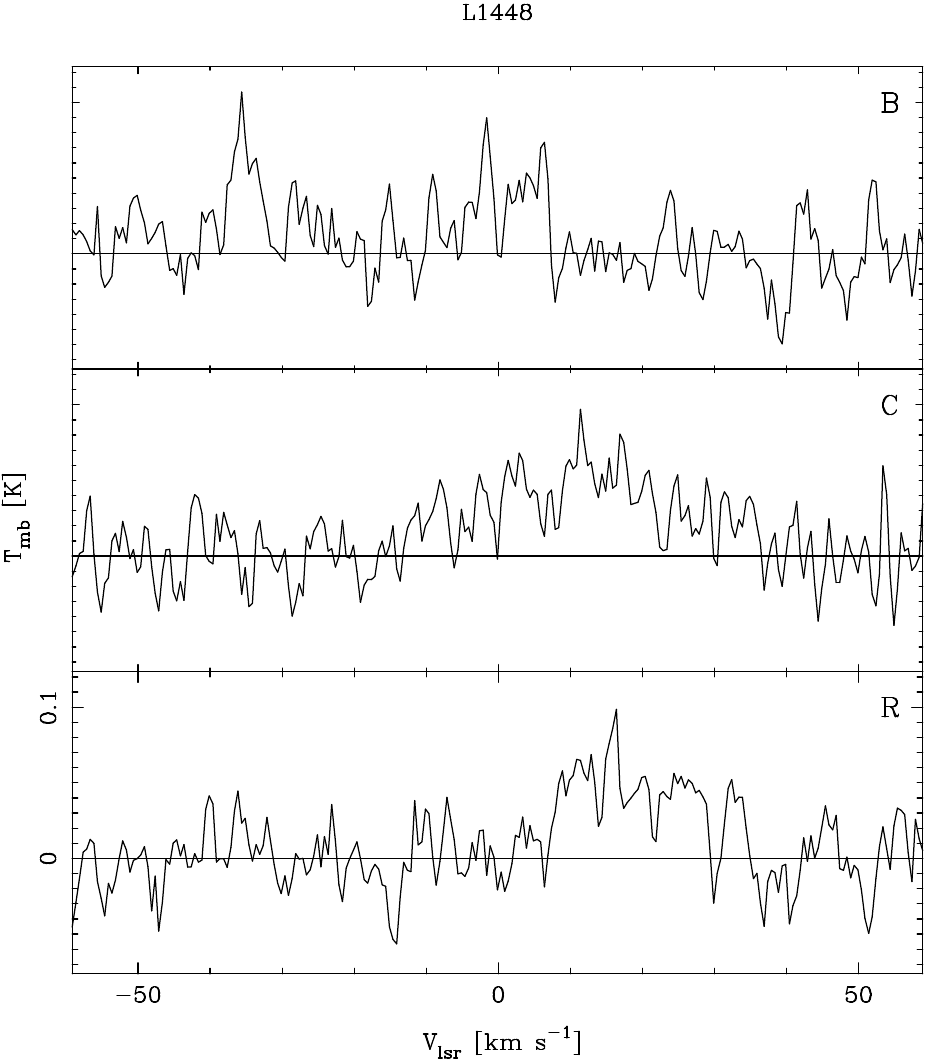}} 
\caption{L1448 spectra. The positions are listed in
  Table~2 and shown in Figure~\ref{fig:L1448}. A letter
  in the upper right corner indicates in which part of the flow the
  spectra have been collected (R=red, B=blue and C=center). The spectra
  have been baseline subtracted and smoothed to a resolution of 0.5
  \kmpers.}
 \label{fig:L1448spectra}
\end{figure}
L1448 is a dark cloud in the constellation of Perseus at the distance
of 250 pc \citep{Enoch:2006lr}. The large, highly collimated outflow
originating from L1448-mm shows enhanced emission from SiO in both
lobes \citep{Nisini:2007fk}. \citet{Franklin:2008fk} report \ohtvao\
abundances of $X$(\ohtvao) = 1.5\texpo{-6} in the blue wing and
$X$(\ohtvao) = 3.7\texpo{-6} in the red wing. We report observations
in three positions across the structure, where the northern position
also covers the outflow from L1448 IRS3. The possible detections in
both lobes can, due to instabilities in the baselines, only be
classified as likely. There is also a tentative detection of a bullet
feature in the northern position at $v_{\mathrm{lsr}} \sim$
+35\,\kmpers and we note that this observation is consistent with the
high speed CO bullet B3 reported by \citet{Bachiller:1990lr}. However,
the preliminary analysis of HCO$^{+}$ data, recently taken at the
Onsala Space Observatory, does not reveal any emission at this
velocity. Mass loss rates of \mdot$_{\rm loss}$ = 4.6\texpo{-6} \msun\
yr$^{-1}$ for L1448-mm and \mdot$_{\rm loss}$ = 1.1\texpo{-6} \msun\
yr$^{-1}$ for L1448 IRS3 were reported by \cite{Ceccarelli:1997kx}
based on CO observations carried out by
  \citep{Bachiller:1990lr}.
$N$(\htva) = 6\texpo{19}\,cm$^{-2}$ and $n$(\htva) =
1\texpo{3}\,cm$^{-3}$ are inferred from mass and size estimates
reported by the same authors. We assume the width of the flow
to be 40\asec.
Taking the gas 
temperature to be $T = 37\,$K for all positions \citep[][dust temperature
towards L1448-mm]{Bachiller:1995uq} we derive \orthowater\ abundances
between 6\texpo{-4} and 2\texpo{-3} in the
outflow. Using the higher volume density ($\sim$\expo{4})
  estimated by \citet{Bachiller:1990lr}, we derive \orthowater\
  abundances between 1\texpo{-4} and 3\texpo{-4}.

\subsubsection{HH211}
The Herbig-Haro jet HH211 is also located in Perseus, 315\,pc away
\citep{Herbig:1998fk}. It was observed in three different positions
enclosing the relatively small
outflow. 
The central and northern beams contain the HH211-mm region. We use the
mass estimates from \citet{Gueth:1999vn} 
as the base for our inferred volume densities, $n$(\htva) =
1\texpo{4}\,cm$^{-3}$ and column densities, $N$(\htva) =
4\texpo{19}\,cm$^{-2}$ in all three positions. The size of the outflow
is taken to be 90\asec\ $\times$ 10\asec. The temperature in the
clumps of the IC348 region varies between 12\,K close to the centers
and 20 -- 30\,K at the edges \citep{Bachiller:1987lr}. Assuming a
temperature $T = 12\,$K we derive upper limits of $X$(\ohtvao) $<$
8\texpo{-4} at the central position and $X$(\ohtvao) $<$ (1 --
2)\texpo{-3} in the outflow. A temperature, $T = 30\,$K, would lower
these upper limits by almost a factor of ten. The two-sided mass loss
rate was estimated by \citet{Lee:2007kx} to be \mdot$_{\rm loss}$
$\sim$ (0.7 -- 2.8)\texpo{-6} \msun\ yr$^{-1}$.

\subsubsection{L1551}
L1551 is probably one of the most rigorously studied molecular
outflows. The main source L1551 IRS5 is located at the distance of
140\,pc in the Taurus-Auriga cloud complex \citep{Kenyon:1994uq}. The
mass loss rate is in the range 8\texpo{-7} $<$ \mdot$_{\rm loss}$ $<$
2\texpo{-6} \msun\ yr$^{-1}$ \citep[][and references
therein]{Liseau:2005lr}.
In this paper we use the mass estimate from CO observations
provided by \cite{Stojimirovic:2006fk} to calculate the
volume density and column density to $n$(\htva) =
3\texpo{3}\,cm$^{-3}$ and $N$(\htva) =
  1\texpo{21}\,cm$^{-2}$, respectively. These authors estimate the
mass of the outflow to be 7.2 \msun\ and we estimate
  the size of the outflow to 1.3 $\times$ 0.2 pc. Assuming a kinetic
temperature of 20\,K we give upper limits on the \orthowater\ abundance in
three positions across the outflow. We obtain upper limits, ranging
from $X$(\ohtvao) $<$ 8\texpo{-5} to $X$(\ohtvao) $<$
  2\texpo{-4}.

\subsubsection{TW Hya}
At the distance of 56\,pc \citep{Qi:2008lr}, TW\,Hya is the most
nearby known T\,Tauri star.  Its accretion disk of size 7\asec\ is
seen essentially face-on and is point-like to \odin, i.e. particularly
for any \htvao\ emission region, the beam filling factor would
presumably be $\ll $ 3\texpo{-3}. In our sample, the object has
the lowest mass loss rate, viz. $10^{-12\,{\rm to}\,-11} \le $
\mdot$_{\rm loss} < $7\texpo{-10}\,\msunyr\
\citep[][respectively]{Dupree:2005fk, Lamzin:2004qy}, and with an age
of about 10\,Myr \citep{de-la-Reza:2006uq}, it is likely also the
oldest. \citet{Herczeg:2004fj} report the detection of
L$\alpha$-excited \htva, with an $N$(\htva) $= 3 \times
10^{18}$\,cm$^{-2}$, which appears accurate to within an order of
magnitude. This value refers to the innermost disk regions. In the
outer disk, CO and other trace molecules, including their isotopic
and/or deuterated forms, have also been detected
\citep{Kastner:1997kx, van-Zadelhoff:2001yq, Qi:2008lr}. In the HCN
forming regions, the densities are determined to be
\expo{6} -- \expo{8}\,cm$^{-3}$ \citep{van-Zadelhoff:2001yq}. From
intermediate $J$-transitions of CO, these authors estimated
temperatures to be in the range $40 \le T < 150$\,K.

Not entirely unexpected, \odin\ did not detect the \htvao\ 557\,GHz
line\footnote{Taking the difference in beam sizes into account, the
  \odin\ data are only a slight improvement over those obtained with
  \swas\ (E.\,Bergin and R.\,Plume, private
  communication).}. For representative disk parameters and a line
width of $< 1$\,\kmpers, the rms of 14\,mK would imply an abundance,
$X$(\ohtvao) $<$ 1\texpo{-8}. %($1\sigma$).
For the modeling we have used a temperature of 40\,K. However,
increasing this parameter to 150\,K will not decrease the derived
upper limit by more than 20\%.

\subsubsection{$\epsilon$ Cha\,{\sc I\,N}}
The star forming cloud Chamaeleon\,{\sc I} is located at a distance of
150\,pc \citep{Knude:1998qy}.  From the estimated age (3.8\texpo{4}
yr), mass (0.21 \msun) and maximum velocity ($\sim$6 \kmpers) reported
by \cite{Mattila:1989lr}, we obtain a mass loss rate, \mdot$_{\rm
  loss}$ = 3.3\texpo{-7} \msun\ yr$^{-1}$. The velocity of the wind is
assumed to be 100 \kmpers.  We estimate $N$(\htva) =
  4\texpo{20}\,cm$^{-2}$ and $n$(\htva) =
  2\texpo{3}\,cm$^{-3}$ from CO observations carried out by
the same authors. The width of the flow is approximately 0.1
  pc. Adopting the temperature 50\,K, given by
  \citet{Henning:1993lr}, we obtain an upper limit,
$X$(\ohtvao) $<$ 3\texpo{-5}.

\subsubsection{Sa136 (BHR71)}
\begin{figure}
  \resizebox{\hsize}{!}{\includegraphics{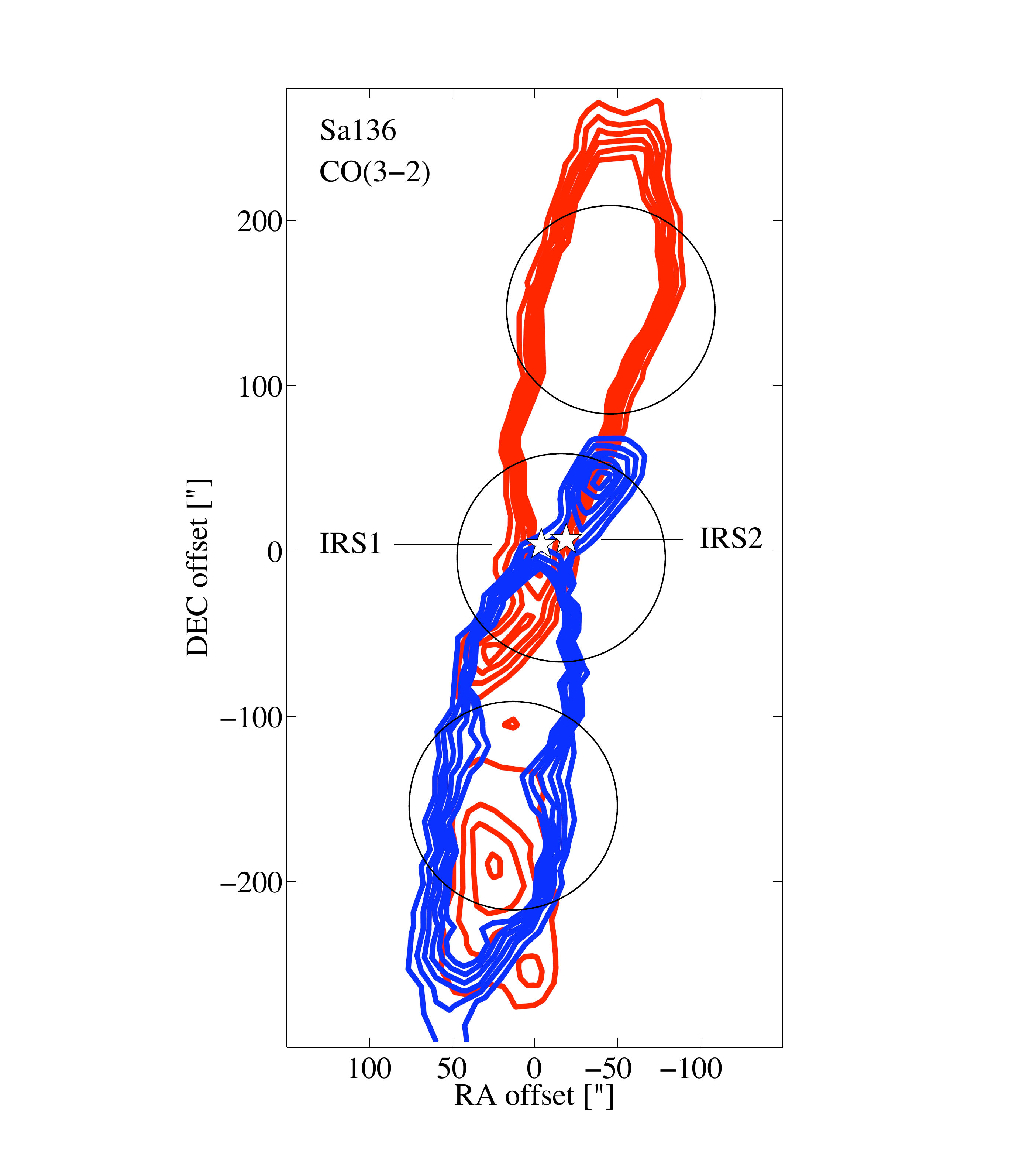}}
  \caption{The three positions observed by \odin\ are shown overlaid
    on a CO (3-2) map of Sa136 \citep{Parise:2006fk}. The circles
    correspond to the Odin beam at 557\,GHz. Coordinate offsets are
    given with respect to: \atwozero = 12:01:37.0, \dtwozero =
    -65:08:53.5. The positions of the sources IRS1 and IRS2 are
    indicated with star symbols.}
  \label{fig:BHR71}
\end{figure} 
\begin{figure}
\resizebox{0.9\hsize}{!}{\includegraphics{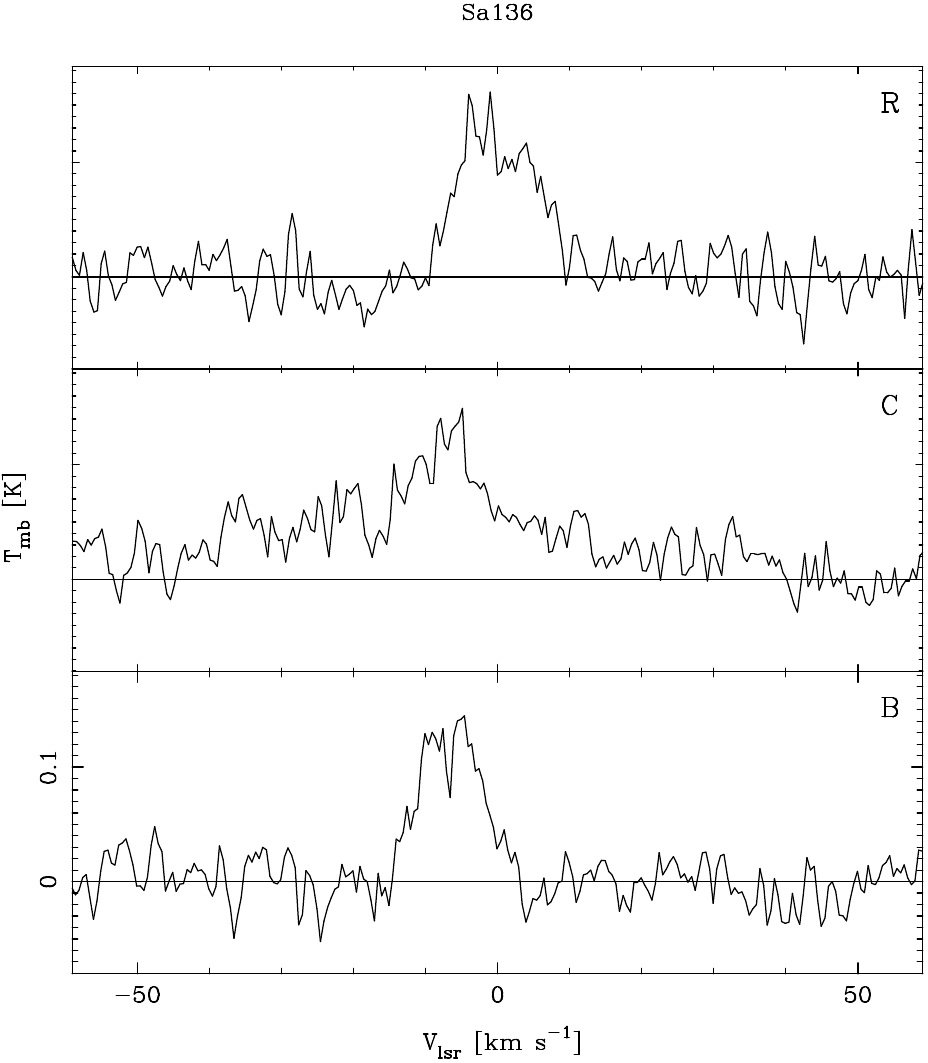}} 
 \caption{The same as Figure~\ref{fig:L1448spectra} but for Sa136.}
 \label{fig:Sa136spectra}
\end{figure}
The Sa136 Bok globule outflows \citep{Sandqvist:1977lr}, at about 200\,pc distance, are driven
by a binary protostellar system. The secondary CO outflow driven by
IRS 2 is more compact \citep{Bourke:2001kx} than the larger outflow
driven by IRS 1.

Taking the column density measurements made by \citet{Parise:2006fk},
assuming a [CO/\htva] ratio of $\rm{10^{-4}}$ we obtain $N$(\htva) =
1\texpo{21}\,cm$^{-2}$, $N$(\htva) = 4\texpo{20}\,cm$^{-2}$ and
$N$(\htva) = 3\texpo{21}\,cm$^{-2}$ in the red, central and blue
part of the flows respectively. We assume that the
  flow has a depth of 0.07 pc, yielding volume densities
  $n(\mathrm{H_2}) =$ 6\texpo{3} cm$^{-3}$, $n(\mathrm{H_2}) =$
  2\texpo{3} cm$^{-3}$ and $n(\mathrm{H_2}) =$ 1\texpo{4}cm$^{-3}$ in
  the same regions. \citet{Parise:2008rt} estimate $T =
30$ -- $50\,$K from CO and methanol observations. In our modeling we use
$T =
40\,$K. 
We estimate the \orthowater\ abundances to (0.1 -- 1)\texpo{-5} in the
outflow and 2\texpo{-4} at the central
position. However, \cite{Parise:2008rt} give a density of
  $n(\mathrm{H_2}) =$ 1\texpo{5} cm$^{-3}$ in the region. Using this
  higher value we obtain \orthowater\ abundances of (2 -- 6)\texpo{-7}
  in the outflow and 3\texpo{-6} towards the central source. The
emission has broader wings in the central position, a feature present
also in the \swas\ data. The origin of this high velocity component
and the elevated water abundance might be the smaller outflow
originating from IRS~2, visible in Figure~\ref{fig:BHR71}. Based on
the outflow mass (1.3 \msun), dynamical time scale (1\texpo{4} yr) and
flow velocity (28 \kmpers) provided by
\citet{Bourke:1997uq}\footnote{These authors estimate $t\rm{_d}$ to 10
  700 years and 10 200 years for the south east and the north west
  lobe respectively.} for the larger flows, we estimate the mass loss
rate to be \mdot$_{\rm loss}$ = 3.6\texpo{-5} \msun\ yr$^{-1}$. The
wind velocity is assumed to be 100 \kmpers.

\subsubsection{HH54\,B}

The Herbig-Haro object HH54\,B is situated in the Cha\,{\sc II} cloud
at roughly 200\,pc \citep{Hughes:1992qy}. During the observations, the
correlator suffered from ripple. This had the effect of increasing the
line intensity and a substantial amount of data had to be
abandoned. The signal detected can therefore only be classified as
tentative although we are confident that the data do not show any
systematic variations. Complementary data were obtained in CO(3-2),
CO(2-1), SiO(5-4), SiO(3-2) and SiO(2-1) with the \sest\ telescope
(see Appendix \ref{appendix:CO}) and in CO(5-4) with \odin. The
CO(5-4) data suffer from frequency drift, something that puts an
$\sim$10\,\kmpers\ uncertainty to our velocity scale. However, we do
not believe that this gives large uncertainties on the line strength.

No shock-enhanced emission was detected in any of the observed SiO
transitions. This result seems not easily reconcilable with the
prediction from theoretical C-shock models \citep[see Fig.\,6
of][]{Gusdorf:2008lr}, which appear closely adaptable to the conditions
in HH\,54 \citep{Liseau:1996fk,Neufeld:2006fk}.

Both the CO\,(2-1) and (3-2) lines show an absorption feature at
$+2.4$\,$\mathrm{km \, s^{-1}}$, which corresponds to the LSR-velocity
of the molecular cloud. In addition, a strong blue wing is observed in
all positions, but essentially no redshifted gas, which is in
agreement with the (1-0) observations by \citet{Knee:1992lr}. Both
(2-1) and (3-2) transitions peak at the central map position, i.e. on
HH\,54B itself and their integrated intensities, $\int\!\!T_{\rm
  A}^{\star}\,d\upsilon$, are given in Table~\ref{SEST_obs}.

For the comparison with the ISO-LWS model of \citet{Liseau:1996fk}, we
use the average radiation temperature, approximated by $<\!T\!>
=\eta_{\rm mb}^{-1}\,\int\!\!T_{\rm
  A}^{\star}\,d\upsilon/\int\!d\upsilon$. This yields $<\!T_{21}\!> =
(1.2 \pm 0.02)$\,K and $<\!T_{32}\!> = (3.2 \pm 0.33)$\,K for the
CO\,(2-1) and (3-2) lines, respectively. Both values are smaller, by
25\% and 40\% respectively, than the model predictions of 1.6\,K and
5.7\,K, which were based on a single-temperature approximation. Even
though the strength of the CO(5-4) line is uncertain, it shows a
slightly higher temperature than the predicted 0.3\,K. Putting it all
together, the model predicts the radiation temperature in all three
lines to within an order of 2. For the \radex\ analysis we use the
ISO-LWS model $T = 330\,$K, $n$(\htva) = 2\texpo{5}\,cm$^{-3}$ and
$N$(\htva) = 3\texpo{19}\,cm$^{-2}$, where the column density has been
diluted to the \odin\ beam. We obtain a beam averaged water abundance
of $X$(\ohtvao) = 3\texpo{-6}. However, recently
\citet{Neufeld:2006fk} estimate a higher \htva\ column density for the
warm gas, a fact that could alter our derived abundance by an order of
magnitude downwards. The mass loss rate of the unknown driving source
has been estimated by \citet{Giannini:2006lr} as \mdot$_{\rm loss} = 3
\times 10^{-6}$ \msun\ yr$^{-1}$.

\begin{figure}
  \resizebox{0.9\hsize}{!}{\includegraphics{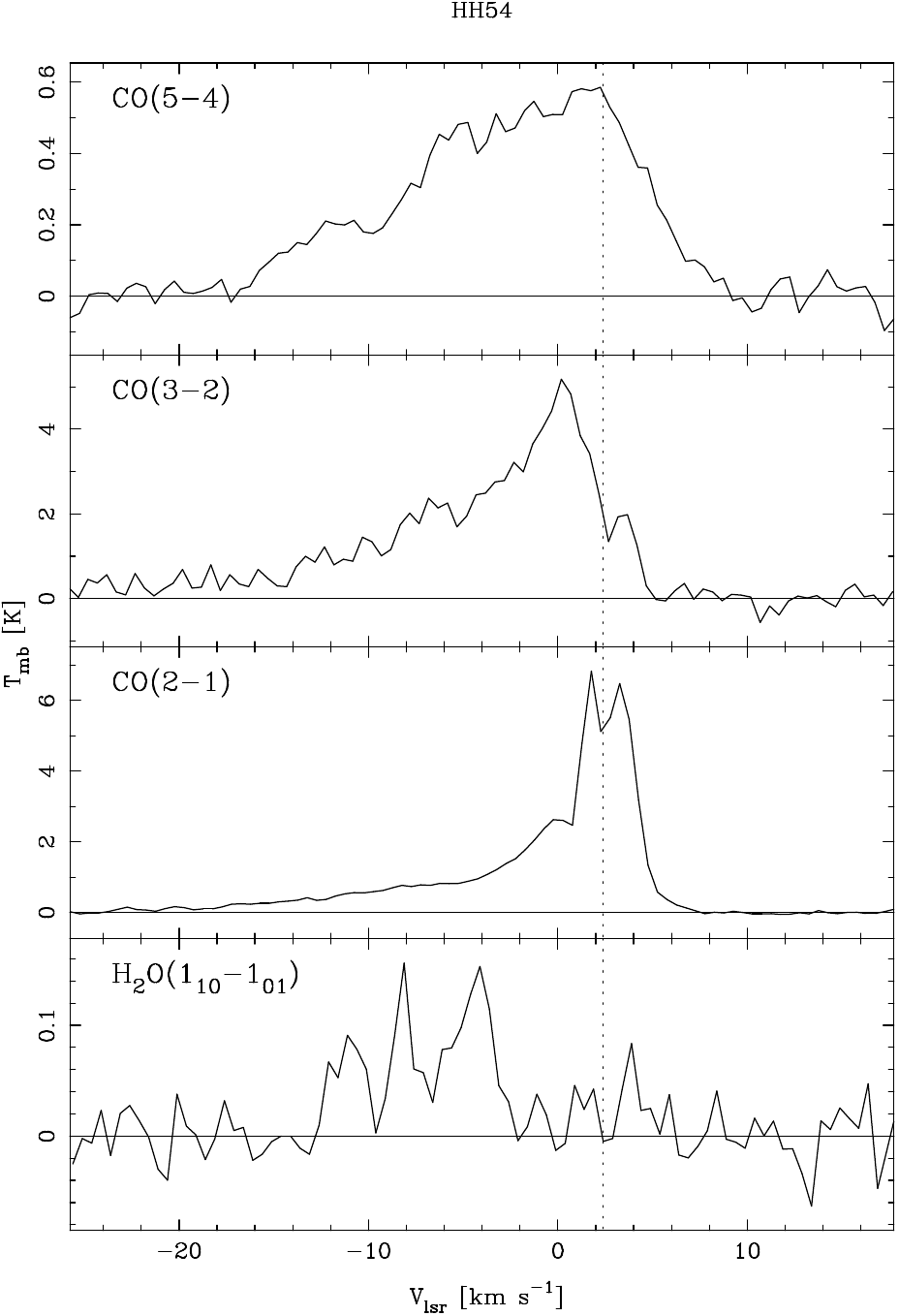}}
  \caption{From top to bottom the CO(5-4), CO(3-2), CO(2-1) and
    $\mathrm{H_2O(1_{10}-1_{01})}$ spectra observed towards HH54\,B
    are plotted. The dotted vertical line shows the cloud LSR-velocity
    at +2.4\,\kmpers.}
  \label{fig:OdinvsSEST}
\end{figure}

\subsubsection{G327.3-0.6}
The hot core G327.3-0.6 is located in the southern hemisphere at the
distance of 2.9\,kpc \citep{Bergman:1992kx}. CO line profiles obtained
by \citet{Wyrowski:2006fj} weakly indicate the presence of outflows,
however, to date there is no further study of this. From CO
observations performed by these authors, we infer $N$(\htva) =
2\texpo{22}\,cm$^{-2}$ and $n$(\htva) = 4\texpo{5}\,cm$^{-3}$, assuming
[$\rm{{}^{18}CO/H_2}$] = \expo{-7} and a source size
  25\asec. The volume density inferred is slightly lower
  than the range \expo{6}--\expo{8}\,cm$^{-3}$ given by
\cite{Bergman:1992kx}, who also estimates the temperature to be within
the range of 40 -- 200\,K. Using a value of 100\,K gives an upper limit
of $X$(\ohtvao) $<$ 8\texpo{-10}. A temperature of 40\,K will increase
the inferred upper limit by a factor of 8.

\subsubsection{NGC6334\,{\sc I}}
\label{ngc6334}
At least two outflows are emerging from NGC6334\,{\sc I}, located in
the constellation Scorpius \citep{McCutcheon:2000yq} at the distance
of 1.7\,kpc \citep{Neckel:1978fj}. From CO observations provided by
\citet{Leurini:2006uq} we obtain a beam averaged column density
1\texpo{20}\,cm$^{-2}$ and a volume density 4\texpo{3}\,cm$^{-3}$. We
assume that the gas temperature is the same as the dust temperature,
viz. T = 100\,K \citep{Sandell:2000fk}. This is consistent with
\citet{Leurini:2006uq} who set a lower limit on the kinetic
temperature at 50\,K. With the above properties we derive an abundance
of $X$(\ohtvao) = 5\texpo{-5}.

The baseline subtracted spectrum does not show any evidence of high
velocity gas. However, we do not find this easily reconcilable with
the high velocity gas detected in several CO transitions by
\citet{Leurini:2006uq}. One possibility could be the curved baseline
hiding the outflow wings. Therefore, we investigate also an
alternative case where we assume that the entire curvature stems from
the outflowing gas. This secondary scenario is perhaps not very
likely. However, at present it is not possible to draw any firm
conclusions. The estimated depth of the absorption feature is greater
than the continuum level of 360 Jy, interpolated from 800 $\mu$m
observations provided by \citet{Sandell:2000fk}. In this secondary
case we derive the abundance $X$(\ohtvao) = 2\texpo{-3}.

\begin{figure}
 \resizebox{0.9\hsize}{!}{\includegraphics{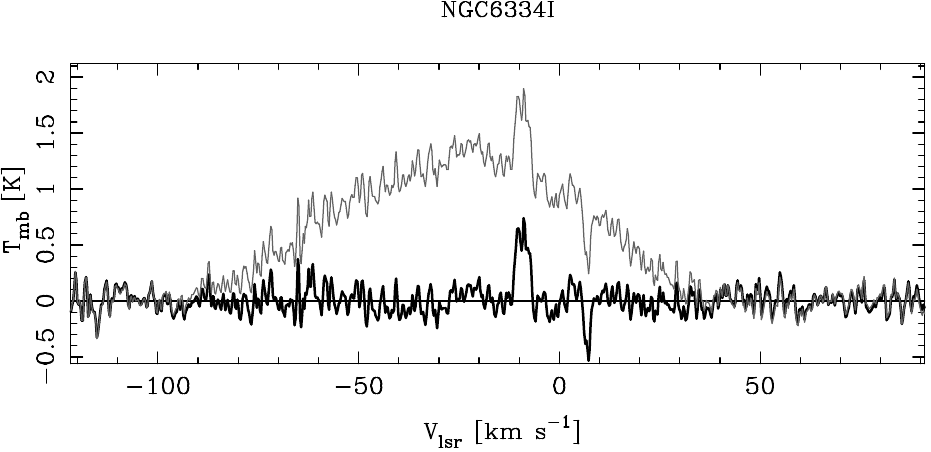}} 

 \caption{The same as Figure~\ref{fig:L1448spectra} but for NGC6334\,{\sc{I}}. The black spectrum represents the first case and the grey spectrum represents the second case as described in Section~\ref{ngc6334}}
 \label{fig:NGC6334Ispectra}
\end{figure}

\subsubsection{Ser SMM1}

The Serpens star forming dark cloud is situated in the inner Galaxy at
a distance of 310\,pc \citep{de-Lara:1991qy}. \cite{Franklin:2008fk}
estimated \orthowater\ abundances of $X$(\ohtvao) = 7.1\texpo{-7} and
$X$(\ohtvao) = 3.8\texpo{-7} in the blue and red wing respectively
while \citet{Larsson:2002lr} estimate the water abundance to
$X$(H$_2$O) = 1\texpo{-5} in the region. \cite{Davis:1999fj} provide
the mass and size of the outflow based on CO observations. Assuming a
width of 0.2\,pc yields $n$(\htva) = 1\texpo{3}\,cm$^{-3}$ and
$N$(\htva) = 5\texpo{20}\,cm$^{-2}$.  From O{\tiny{I}}(63$\mu$m)
measurements carried out by \citet[][and references
therein]{Larsson:2002lr} we obtain the mass loss rate, \mdot$_{\rm
  loss}$ = 3\texpo{-7} \msun\ yr$^{-1}$. The temperature of the dust
was constrained by \citet{White:1995lr} to be 30\,K $< T <$
40\,K. Using $T = 35\,$K and the above properties we obtain
$X$(\ohtvao) = 9\texpo{-5} and $X$(\ohtvao) = 5\texpo{-5} in the blue
and red flow.

\begin{figure}
\resizebox{0.9\hsize}{!}{\includegraphics{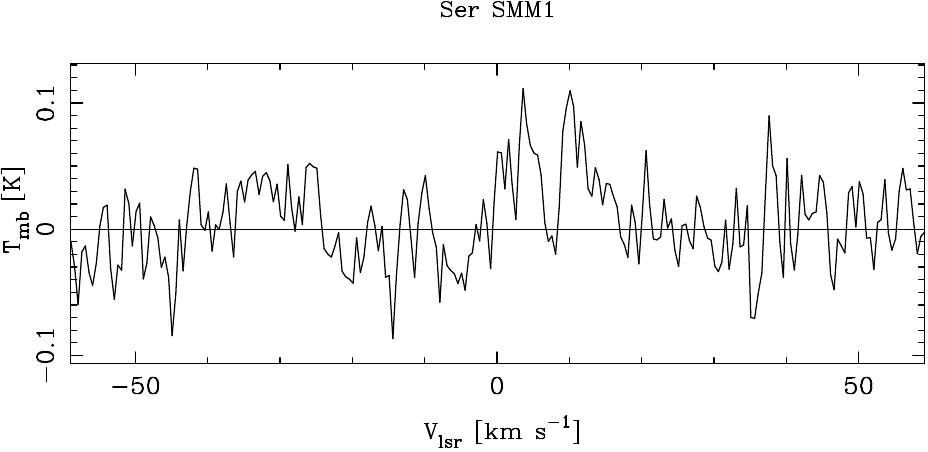}} 
 \caption{The same as Figure~\ref{fig:L1448spectra} but for Ser SMM1.}
 \label{fig:SerSMM1spectra}
\end{figure}

\subsubsection{B335}

The dense core in the B335 globule is believed to be one of the major
candidates for protostellar collapse. This isolated source at the
distance of 150\,pc \citep{Stutz:2008fk} harbors several Herbig-Haro
objects associated with a bipolar outflow \citep[see
e.g.][]{Galfalk:2007lr}. Following the outflow mass estimate of 0.44
\msun\ in \cite{Hirano:1988lr} we derive $n$(\htva) =
1\texpo{3}\,cm$^{-3}$ and $N$(\htva) = 3\texpo{20}\,cm$^{-2}$. The
size of the outflow is taken to be 2\amin\ $\times$ 8\amin.  Based on
the same mass, a dynamical time scale (2.3\texpo{4} yr) and a flow
velocity (13 \kmpers), provided by these authors we derive \mdot$_{\rm
  loss}$ = 2.5\texpo{-6} \msun. The velocity of the wind is assumed to
be 100 \kmpers. In our modeling we use a kinetic temperature of
20\,K. This is an intermediate value in the modeling carried out by
\citet[][their Figure 8]{Evans:2005qy} who give the temperature as a
function of radius for the inner 0.1\,pc region. The derived upper
limit of the water abundance is, $X(o$-$\mathrm{H_2O}) < $
1\texpo{-3}. The rms implies a water abundance, $X(o$-$\mathrm{H_2O})
< \rm{\sim 10^{-7}}$ according to the infall model presented by
\citet{Hartstein:1998fk}. \citet{Ashby:2000fk} estimate the upper
limit on the \orthowater\ abundance to be within the range
1.3\texpo{-7} -- 7\texpo{-6}.
\subsubsection{L1157}
\begin{figure}
  \resizebox{0.9\hsize}{!}{\includegraphics{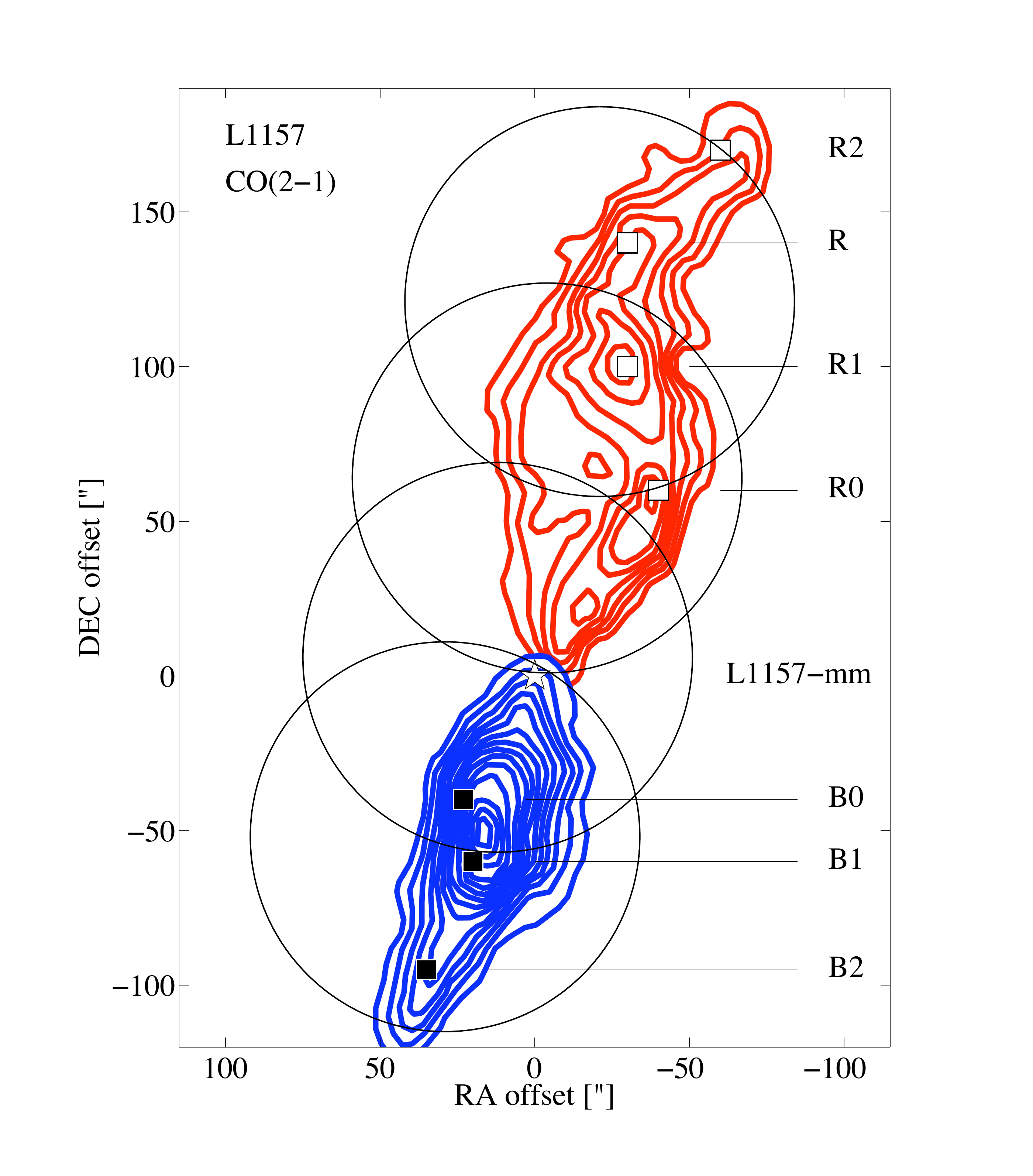}}
  \caption{The four positions observed by \odin\ are shown overlaid on
    a CO (2-1) map of L1157 \citep{Bachiller:2001lr}. The circles
    correspond to the Odin beam at 557\,GHz. Coordinate offsets are
    given with respect to L1157-mm, indicated in the figure with a
    star symbol: \atwozero = 20:39:06.4, \dtwozero = +68:02:13.0. The
    black and white squares refer to the knots B0, B1, B2, R0, R1, R
    and R2 described in \citet{Bachiller:2001lr}.}
  \label{fig:L1157}
\end{figure}  

\begin{figure}
\resizebox{0.9\hsize}{!}{\includegraphics{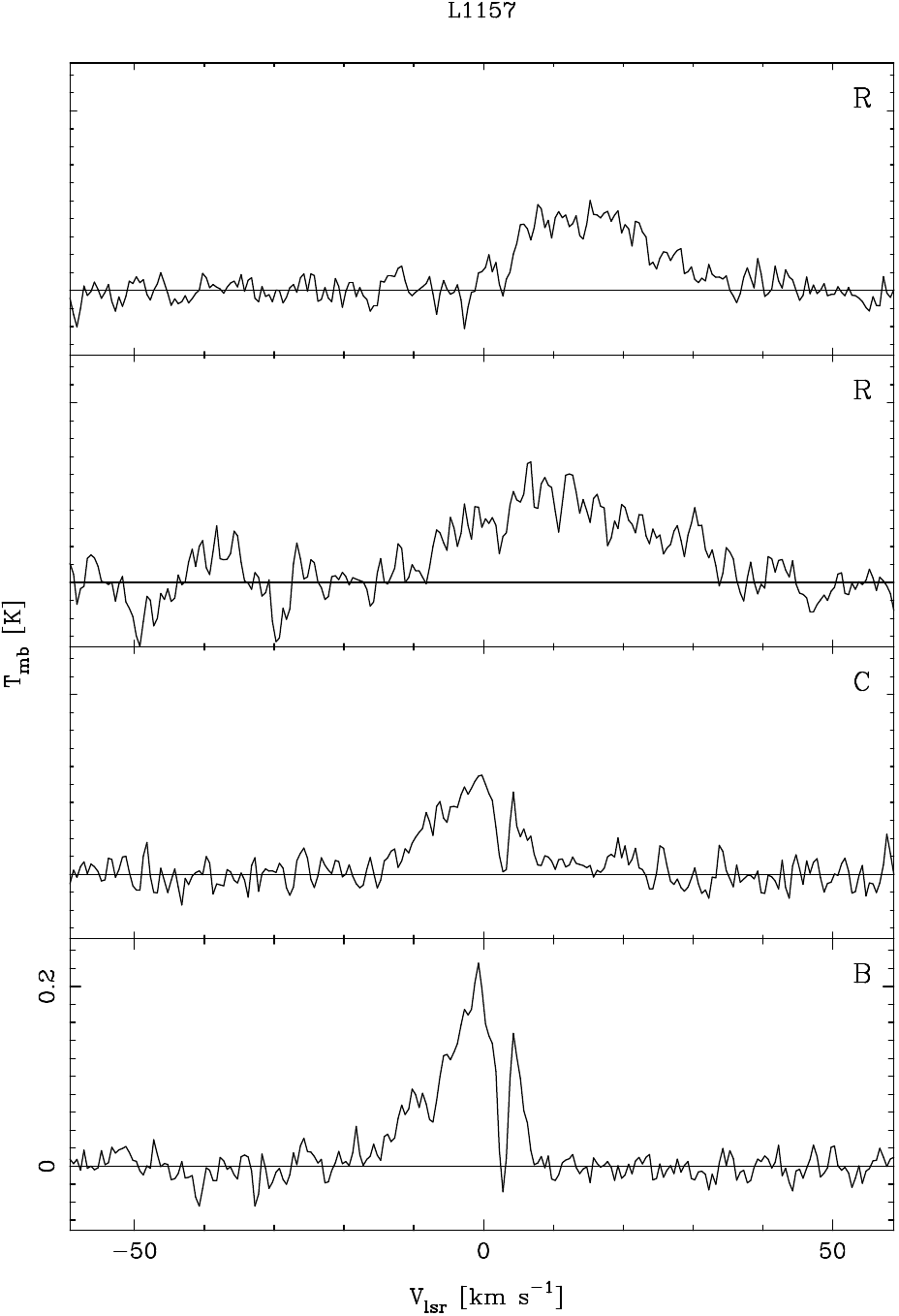}} 
 \caption{The same as Figure~\ref{fig:L1448spectra} but for L1157.}
 \label{fig:L1157spectra}
\end{figure}

L1157 is a Class 0 object in the constellation of Cepheus situated at
the distance of 250\,pc \citep{Looney:2007lr}.
It drives a prototype bipolar outflow which was observed by \odin\ in
four different positions. The mass loss rate can be derived from mass
(0.62 \msun) and time scale (15 000 yr) estimations given by
\citet{Bachiller:2001lr}. Assuming a maximum CO velocity of 20
\kmpers\ and a stellar wind velocity of 100 km s$^{-1}$, we derive
\mdot$_{\rm loss}$ = 8.3\texpo{-6} \msun\
yr$^{-1}$
. The \odin\ strip map covers the bulk of the outflow with two
pointings in the red wing, one in the blue, and one on the driving
source of the outflow itself (Figure~\ref{fig:L1157}). The
observations were carried out using the AOS and the AC2
simultaneously.
The spectra shown in Figure~\ref{fig:L1157spectra} are the averages of
the merged data.  The total mass in the different
  parts of the flow were obtained from CO
observations carried out by \cite{Bachiller:2001lr}. From
this we estimate $n$(\htva) = 2\texpo{3}\,cm$^{-3}$ and $N$(\htva) =
2\texpo{20}\,cm$^{-2}$ in the northern lobe, while $n$(\htva)
= 3\texpo{3}\,cm$^{-3}$ and $N$(\htva) = 2\texpo{20}\,cm$^{-2}$ in the
central and southern region. The size of the CO outflow is
  taken to be 50\asec\ $\times$ 375\asec.
  For all four positions we set the kinetic temperature to $T =
  30\,$K, a rough global estimate based on \citet{Bachiller:2001lr}.
  We calculate the abundances in the outflow to be within the range of
  $\rm{2 \times 10^{-4}}$ and $\rm{1 \times 10^{-3}}$. The derived
  water abundance in the central region is slightly lower,
  $X$(\ohtvao) = 2\texpo{-4} in the blue lobe and $X$(\ohtvao) =
  3\texpo{-5} in the red. The increased blue emission likely
  originates from the outflow. Within the \odin\ beam are the
  positions B0 and B1 that show peaked emission in H$_2$CO, CS,
  CH$_3$OH, SO \citep[][their Figure 1]{Bachiller:2001lr} and SiO
  \citep{Nisini:2007fk}. \cite{Franklin:2008fk} estimated $X$(\ohtvao)
  = 8.0\texpo{-6} and $X$(\ohtvao) = 9.7\texpo{-6} in the blue and red
  wing respectively.

\citet{Bachiller:2001lr} estimates the density around the protostar
to be $\sim$\expo{6}\,cm$^{-3}$. When moving from B0 to B2, the
density changes from $\sim$3 to 6\texpo{5}\,cm$^{-3}$. Using
$n$(\htva) = 1\texpo{6}\,cm$^{-3}$ and $n$(\htva) =
5\texpo{5}\,cm$^{-3}$ for the central and southern part respectively
we obtain the abundances $X$(\ohtvao) = 5\texpo{-7} and $X$(\ohtvao) =
2\texpo{-6}.

\subsubsection{NGC7538 IRS1}

NGC7538 IRS1 is a region of ongoing high mass star formation. The main
infrared source IRS1 is located at the boundary of an H{\sc II} region
in the Perseus arm, located at a distance of 2.7\,kpc
\citep{Moscadelli:2008uq}.  In addition to IRS1, and its high velocity
outflow, also several other sub-mm sources fall into the large \odin\
beam. The mass loss rate from IRS1 was estimated by
\cite{Kameya:1989fj} to be \mdot$_{\rm loss}$ = 1\texpo{-4} \msun\
yr$^{-1}$. The total mass and size of this region is given by the same
authors, yielding a volume density and column density of $n$(\htva) =
3\texpo{3}\,cm$^{-3}$ and $N$(\htva) = 4\texpo{20}\,cm$^{-2}$
respectively. The kinetic temperature is assumed to be the same as the
dust temperature, i.e. T = 40 \citep{Sandell:2004qy}. We infer a limit
to the water abundance of $X$(\ohtvao) $<$ 1\texpo{-4}.

\begin{figure}
\resizebox{0.9\hsize}{!}{\includegraphics{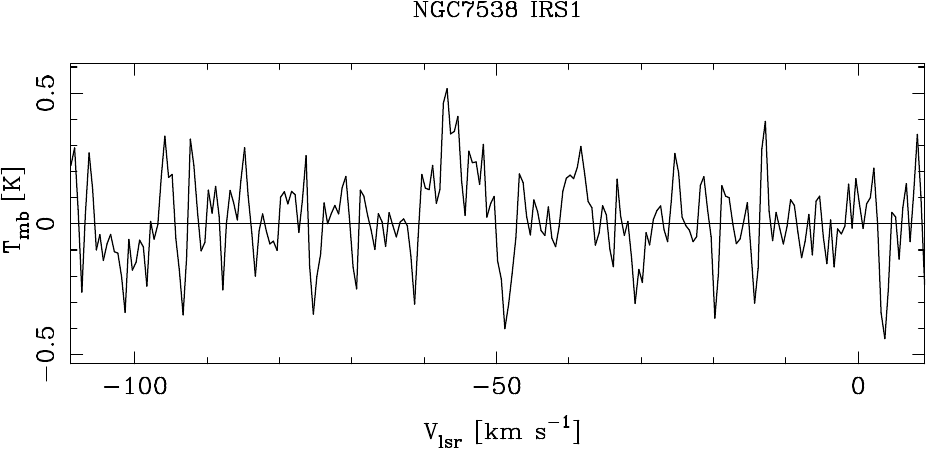}} 
 \caption{The same as Figure~\ref{fig:L1448spectra} but for NGC7538 IRS1.}
 \label{fig:NGC7538spectra}
\end{figure}

\subsubsection{Supernova remnants}
%\begin{description}
\textbf{3C391 BML} is a supernova remnant at the distance of 8\,kpc
\citep{Chen:2001fk}. The temperature, volume density and column
density of the gas were constrained to 50\,K $\leq T \leq$
125\,K, \expo{4}\,cm$^{-3} \leq n$(\htva) $\leq$ 5\texpo{5}\,cm$^{-3}$
and $N$(\htva) $\gtrsim$ 4\texpo{20}\,cm$^{-2}$ from OH 1720
  MHz maser observations carried out by \citet{Lockett:1999qy}. Using
the lowest values for the temperature and volume density we calculate
an upper limit of $X(o$-$\mathrm{H_2O}) <$
4\texpo{-6}.\\\textbf{IC443} is a shell type supernova remnant at the
distance of about 1.5\,kpc \citep{Fesen:1984vn}. The broad emission
peaks are labeled A through H, where \odin\ has observed clump G. The
volume density and temperature were modeled by
\cite{van-Dishoeck:1993fk} as $n$(\htva) = 5\texpo{5}\,cm$^{-3}$ and
$T = 100\,$K.  Using the CO column density inferred by the same
authors, assuming a $\rm{[CO/H_2]}$ ratio of $\rm{10^{-4}}$ and a
source size of 40\arcsec\ $\times$ 100\arcsec, we obtain $N$(\htva) =
3\texpo{21} cm$^{-2}$. The derived abundance is $X(o$-$\mathrm{H_2O})
= \rm{4\times 10^{-8}}$. This is in agreement with
\citet{Snell:2005lr} who derive an \ohtvao\ abundance with respect to
$^{12}$CO, $X$(\ohtvao) = 3.7\texpo{-4}.
\begin{figure}
\resizebox{0.9\hsize}{!}{\includegraphics{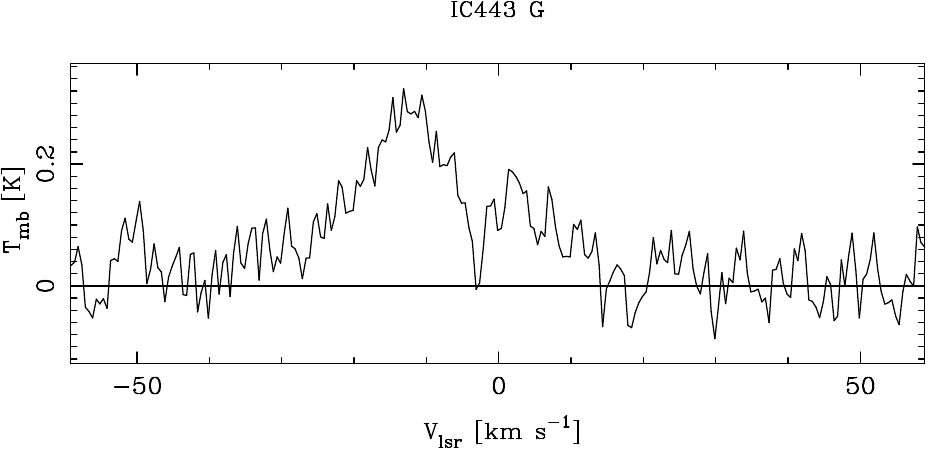}} 
 \caption{The same as Figure~\ref{fig:L1448spectra} but for IC443-G.}
 \label{fig:IC443-Gspectra}
\end{figure}

\subsection{Water abundance}
\begin{figure}
\begin{flushleft}
  \resizebox{0.9\hsize}{!}{\includegraphics{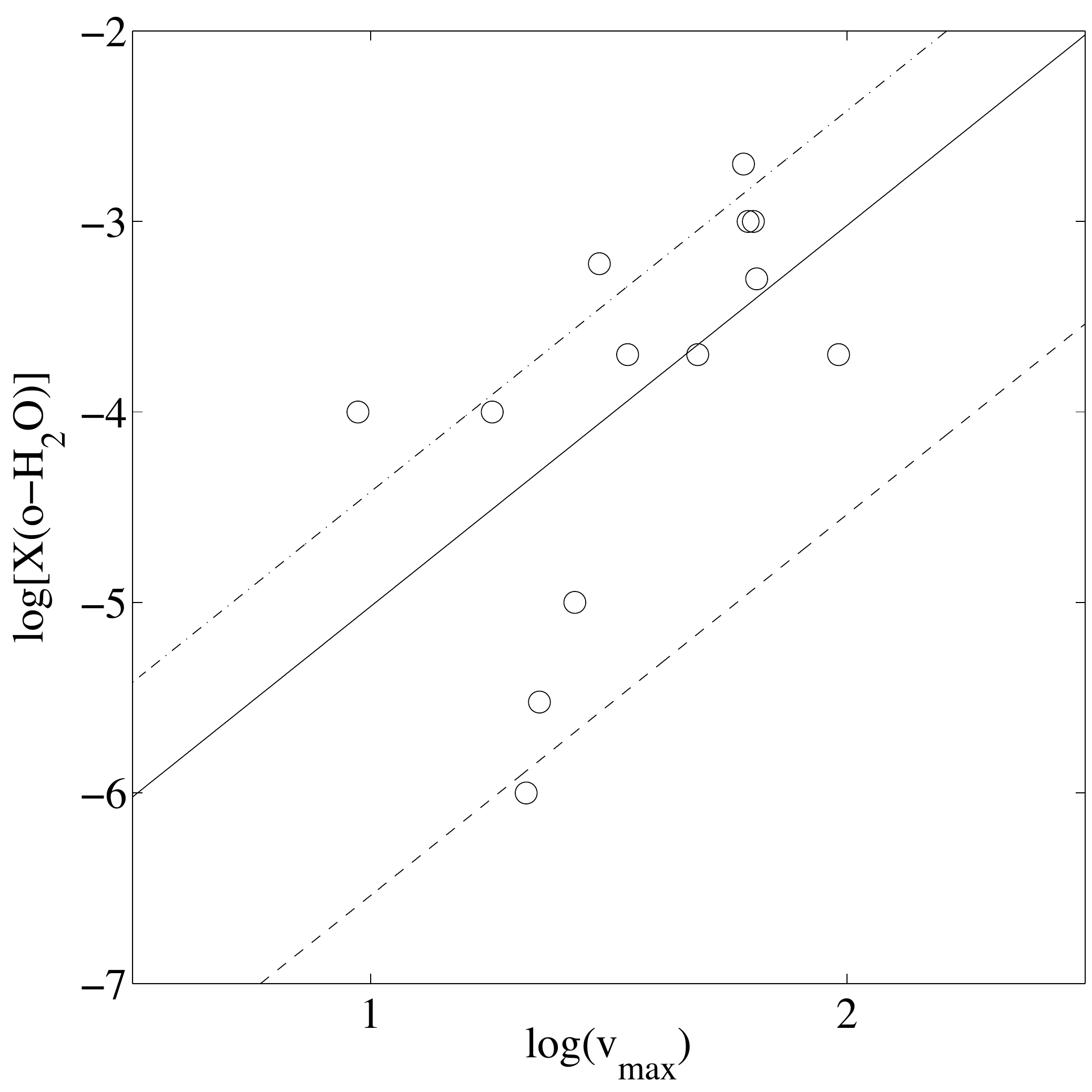}}
  \caption{\ohtvao\ abundance plotted against the maximum velocity for
    an overall inclination of 60$^{\circ}$ with respect to the line of
    sight. The solid line represents a linear fit of
    $\log$[$X$(\ohtvao)] versus $\log (\upsilon_{\rm max})$ with the
    same inclination angle applied. The dotted dashed line represents
    the fit with a inclination correction of 35$^{\circ}$ while the
    dashed line represents the fit with an inclination angle
    correction of 85$^{\circ}$. The errorbars from the measurements are
    smaller than the circles.}
  \label{fig:velocity}
\end{flushleft}
\end{figure}

\begin{figure}
  \resizebox{0.9\hsize}{!}{\includegraphics{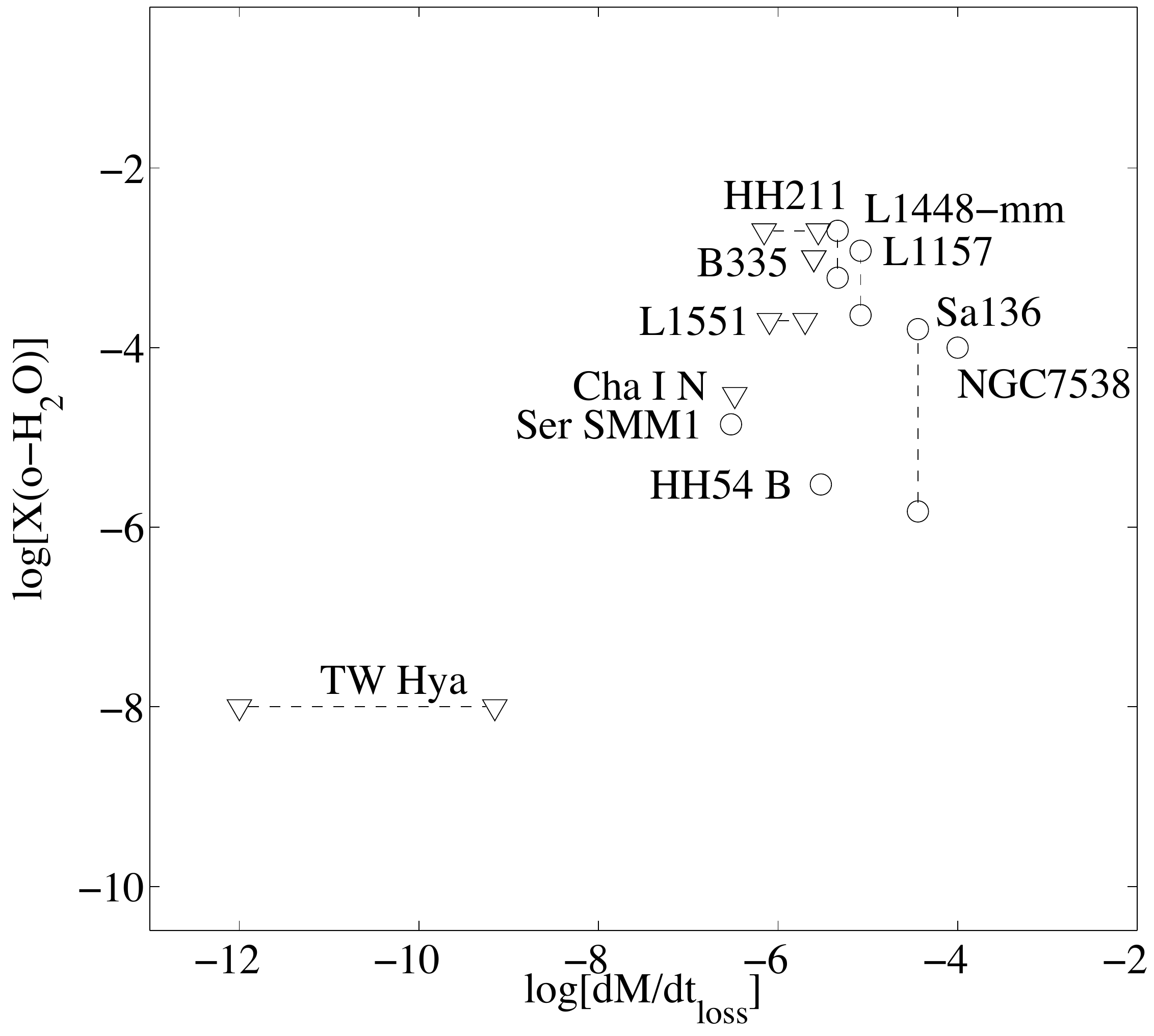}}

  \caption{\ohtvao\ abundance plotted against the mass loss rate. The
    triangles represent the high upper limits for each source on the
    \ohtvao\ abundance, while the circles symbolize values where a
    detection has been made. Dashed lines represent the cases where
    there is a range in the inferred abundances or mass loss
    rates.}
  \label{fig:massloss}
\end{figure}

The water abundances inferred from our analysis are given with respect
to the molecular column density within the \odin\ beam. The values
span a wide range with most of the sources having an abundance of the
order $\rm{10^{-5}}$--$\rm{10^{-4}}$. The highest abundances
in our sample are found in the outflows of L1157 and
  L1448. Supposing that the assumed physical properties are correct,
we see an increased abundance in both the blue and the red wing
of L1157. The derived water abundances are significantly low
in G327.3-0.6. The reason for this is the combination of the high
\htva\ column density and temperature used. The water emission is
also very likely beam diluted due to the large distance, resulting in
lower beam averaged \ohtvao\ abundances.
\\ \\
Assuming an inclination angle of 60$^{\circ}$ with respect to the line
of sight for all the targets we plot the \ohtvao\ abundances versus
the maximum velocities (Figure~\ref{fig:velocity}). The maximum
velocities are taken from the spectra as the maximum offsets between
the cloud velocity and the flow velocity. There is a correlation
between the derived abundances and the maximum velocities of the
outflowing gas. The solid line in Figure~\ref{fig:velocity} is the
first order polynomial least square fit:
\begin{equation}
 X(o \mbox{-} \mathrm{H_2O}) = 10^{-7}\upsilon_{max}^{2}.
\end{equation}
The correlation coefficient is 0.57 while the p-value, testing the
hypothesis of no correlation is 0.04. Following the $C$-shock
modelling carried out by \citet{Kaufman:1996qy}, a relationship is
expected when shocks are responsible for the emission. However, the
abundances do not show any tendency to level off with velocities
larger than $\sim$20 \kmpers\ as shown in their Figure~3. This is an
indication of a lack of $J$-type dissociative shocks in this sample of
outflows. There are however uncertainties in our simplified analysis,
for example the inclination of the outflow lobes and the \radex\
analysis. The dependence between the water abundance and the maximum
velocity is consistent with the analysis made by
\citet{Franklin:2008fk}, although their method is based on binning the
weak outflow lines into intervals of 5\,\kmpers.
We also plot the \orthowater\ abundances versus mass loss rates in
Figure~\ref{fig:massloss}. G.327-0.6 is not included due to the lack
of observations towards the assumed outflow, neither is
NGC6334\,{\sc{I}} included in Figure~\ref{fig:velocity} -
\ref{fig:massloss} due to the curvature in the baseline. No obvious
relationship can be picked out.
The absence of a correlation is surprising inasmuch as
high mass loss rates affect the budget of material available for
water formation.

\subsection{Comparison with \swas\ data}
\label{The gas volume density}
\begin{figure}
  \resizebox{0.9\hsize}{!}{\includegraphics{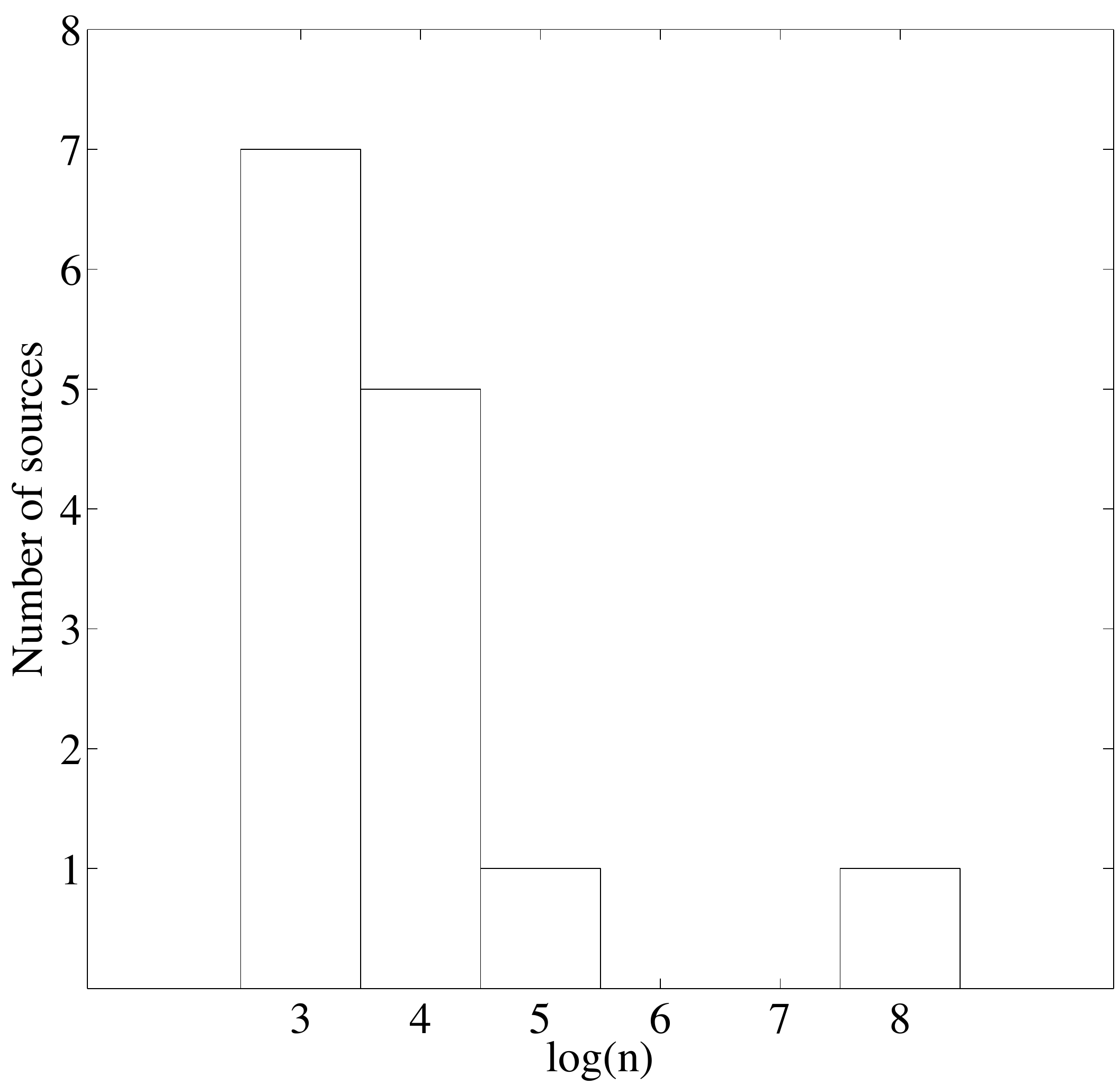}}
  \caption{A histogram of the gas volume density estimated in the outflows
    studied in this paper shows a variation that spans over six
    orders of magnitude.}
  \label{fig:density}
\end{figure}

The gas volume density, used as an input parameter to \radex, has for
all the outflows except TW Hya been derived from CO observations. This
method generally underestimates the mass of the regions. The
possibility that the water emission is originating from gas with a
higher volume density than this can therefore not be ruled
out. Nevertheless, we are confident that the volume density is
actually spanning over a wide range of values. This is also one of the
reasons why several of our derived abundances deviate from those
inferred by \citet{Franklin:2008fk}. They use a single volume density
of $n$(\htva) = \expo{5}\,cm$^{-3}$ and our values differ from this by
more than two orders of magnitude for some of the sources (see
Table~2). Figure~\ref{fig:density} shows a histogram of the numbers of
sources within different volume density ranges. The difficulty to
estimate the gas column density of the water emitting regions is a
problem that has to be adressed in order to interpret the future
observations with Herschel.

\begin{figure}
  \resizebox{0.9\hsize}{!}{\includegraphics{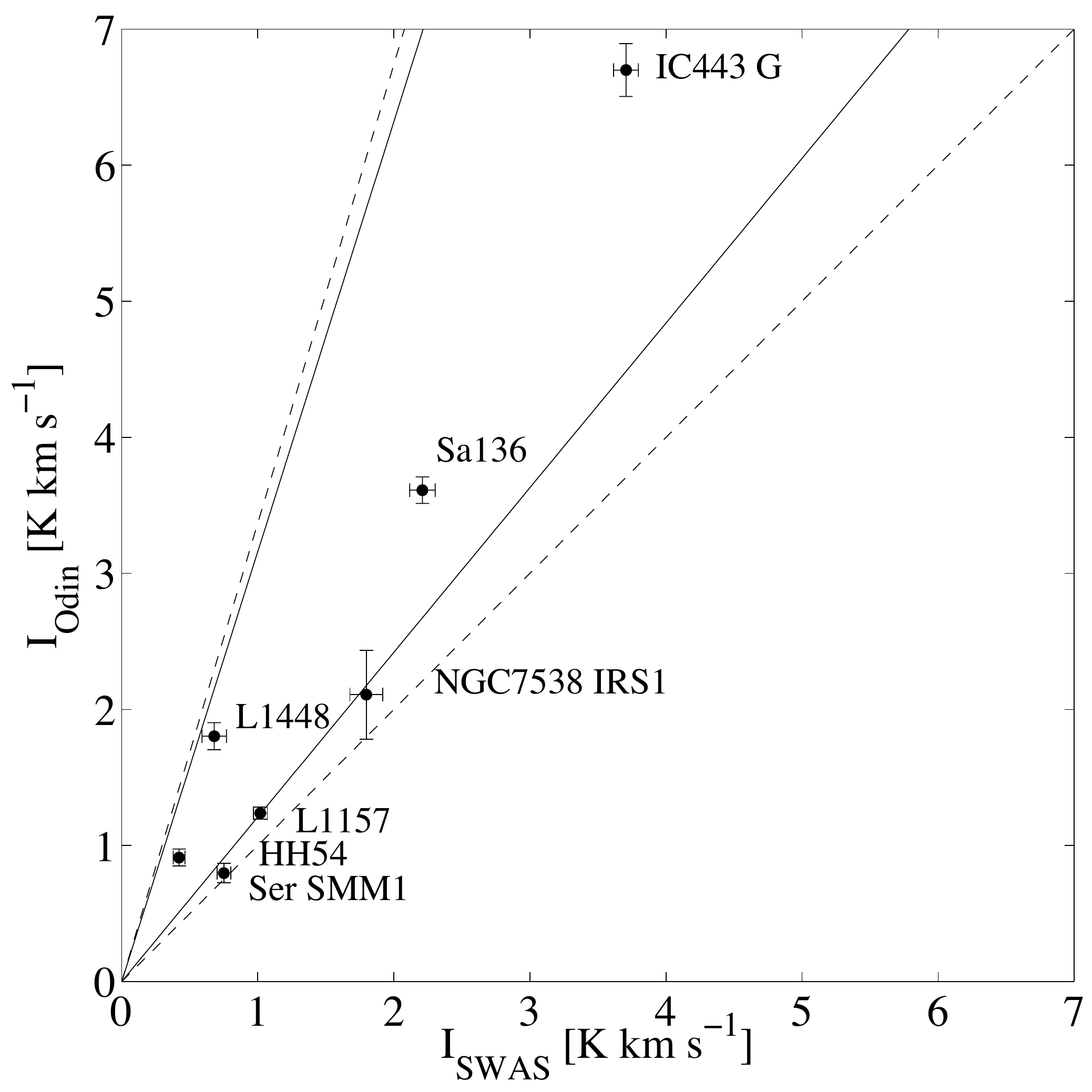}}
  \caption{Integrated intensities for common sources of \odin\ and
    \swas. The dashed lines show the 3.37:1 and the 1:1 ratios between
    the \odin\ and \swas\ integrated intensities. The error bars refer
    to the analysis and the solid lines represents ratios 1:1 and
    3.37:1 with a 15~\% uncertainty applied. This is the estimated
    error limit from the data reduction.}
  \label{fig:OdinvsSWAS}
\end{figure}

The integrated intensities for the \swas\ and \odin\ outflow spectra
are compared in Figure~\ref{fig:OdinvsSWAS} to provide an estimate of
the source size of the water emitting regions. The dashed 1:1 ratio
line illustrates the case where the source is filling both antenna
beams while the dashed 3.37:1 line is indicating a small source size
compared to both beams. Assuming that both the emitting sources and
the antenna responses are circularly symmetric and Gaussian, these
ratios follow from the relation:
\begin{equation}
  \label{eq:3}
  \frac{I_{\tiny{\mathrm{Odin}}}}{I_{\tiny{\mathrm{SWAS}}}} = \frac{(3\amindot\, 3 \times 60 \times 4\amindot\, 5 \times 60) + \theta^2}{126^2 + \theta^2},
\end{equation}
where $\theta$ is source size.  The \swas\ spectra have been retrieved
from the \swas\ spectrum server and baseline subtracted. The
integrated intensities are measured in the same regions as for the
\odin\ spectra and the spectra chosen for comparison are those where
the pointing is equal or within a smaller fraction of a beam. We can
conclude that the sources are filling a large fraction of the beam for
Ser SMM1, L1157 and NGC 7538 1. On the other hand, the source sizes
are small for HH54\,B and L1448.

\subsection{Outflows and observed water abundances}
The main objective for these observations was to use the ground state
\ohtvao\ transition as a tracer for shocked gas. Available shock
models by \citet{Bergin:1998lr} show that a shock velocity in excess
of 10\,\kmpers\ should convert essentially all oxygen, not in
  the form of CO, into water resulting in water abundances relative
to \htva\ of the order $10^{-4}$. This water is also believed to
persist for $\sim$\expo{5} years. For several of our sources,
the water abundance increases as high as what would be theoretically
possible for these $C$-type shocks. The fact that we are observing
abundances as high as this, might be an indication that the shock
velocities are in this regime.
On the other hand, the large \odin\ beam does not show whether the
water emission regions are small or not. The density in these regions
may also very well be orders of magnitude higher than the values used
in this paper, a fact that could alter our inferred values
significantly. Future observations with the \textit{Herschel space
  observatory} will give us the ability to spatially resolve outflows
and it will also give us the possibility to observe several different
transitions originating in gas of different temperatures. This will
hopefully give a much deeper insight into the chemistry occurring in
these structures.

\section{Conclusions}
We make the following primary conclusions:
\begin{enumerate}
\item We have observed 13 outflows and two supernova remnants and detect
  the \orthowater\ ground state rotational transition in seven outflows
  and one super-nova remnant.
\item The column densities of \ohtvao\ have been investigated with
  \radex, having the volume density, temperature, line intensity and
  linewidth as input. Elevated abundances of water are found in
  several sources. The abundances are as high as one
  would expect if all gaseous oxygen had been converted to water in a
  C-type shock.
\item There is no distinct relationship between the water abundance and
  the mass loss rate.
\item There is a correlation between the \ohtvao\ abundance and
  the maximum velocity of the gas.
\end{enumerate}
${}$\\
\textit{Acknowledgments.}  The author enjoyed interesting discussions
with John H. Black concerning radiative transfer in general and
\radex\ in particular. Carina M. Persson and Per Bergman are also
greatly acknowledged. We thank the Research Councils and Space
Agencies in Sweden, Canada, Finland and France for their financial
support. The valuable comments made by the anonymous referee are highly
appreciated.

\longtabL{2}{
\begin{landscape}
\label{tab:table2}
\begin{longtable}{l l r r r r r r r r r r r}
   \caption{Column densities of \ohtvao\ and estimates of the \orthowater\ abundance, $X$(\ohtvao) = $N$(\ohtvao)/$N$(\htva). The offsets are the reconstructed pointing offsets and the numbers in the parentheses are the 1 $\sigma$ statistical uncertainties. Upper limits are 3 $\sigma$.}   \\           % title of Table
\hline
\hline                      
\noalign{\smallskip} 
&&&&& &&Method 1&&&&Method 2&\\
\cline{7-9}\cline{11-13}
\noalign{\smallskip}
Source & Offset & $\int T_{\mathrm{mb}}$(H${}_2$O)$d \upsilon$ &  $\Delta \upsilon$  & $N\mathrm{(H_2)}$ & $T_{\rm kin}$ & $n\mathrm{(H_2)}$& $N$(\ohtvao)& $X$(\ohtvao) &&$n\mathrm{(H_2)}$&$N$(\ohtvao)&$X$(\ohtvao)\\
     & (\asec,\asec) & (K km s${}^{-1}$) & (km s${}^{-1}$)  & (\expo{19} cm${}^{-2}$) & (K) & (\expo{3} cm${}^{-3}$)& (\expo{15} $\mathrm{cm^{-2})}$ & (\expo{-5})&&(\expo{3} cm${}^{-3}$)&(\expo{15} $\mathrm{cm^{-2})}$&(\expo{-5})\\
 \noalign{\smallskip}
     \hline
 \noalign{\smallskip}
 \noalign{\smallskip}
 \multicolumn{3}{l}{\textbf{Odin sources analyzed in this paper:}} \\
\noalign{\smallskip}
 L1448    & $(-38,\,+69)$ & 0.53 (0.63) & $-$10 $\rightarrow$ 7  & 6 & 37 & 1 & \hspace{0.28cm}37 & \hspace{0.28cm}60&& 10&\hspace{0.28cm}7&10\\
          & $(+2,\,-2)$   & 1.80 (0.10) & $-8$ $\rightarrow$ 29  & 6 & 37 & 1& \hspace{0.28cm}91 &\hspace{0.28cm}100 &&10&\hspace{0.28cm}19&30\\
          & $(+32,\,-81)$ & 1.19 (0.07) & 8 $\rightarrow$ 35  &6 & 37 & 1 & \hspace{0.28cm}102 &\hspace{0.28cm}200 &&10&\hspace{0.28cm}20&30\\ 

 HH 211 & $(-47,\,+4)$     &  $<$ 1.01 &   &4 & 12 & 10& $<$ 56 & $<$ 200 &&&&\\
        & $(+14,\,-21)$    &  $<$ 0.56 &   &4 & 12 & 10& $<$ 30 & $<$ 80 &&&&\\
        & $(+73,\,-46)$    &  $<$ 0.96 &   &4 & 12 & 10& $<$ 53 & $<$ 100 &&&&\\

 L1551 & $(+84,\,+18)$    &  $<$ 1.29      &  & 100 & 20 & 3 & $<$ 79 &$<$ 8&& & & \\
            & $(-129,\,-194)$ &  $<$ 1.43 &   & 100 & 20 & 3 & $<$ 88 &$<$ 9&& & & \\
            & $(-342,\,-404)$ &  $<$ 2.36 &  & 100 & 20 & 3 & $<$ 159 &$<$ 20&& & & \\ 
     
 IC443-G  & $(-2,\,-28)$  & 6.70 (0.19) & $-$39 $\rightarrow$ 15 &300&100& 500 &\hspace{0.28cm}0.09 & 0.004&&&& \\
  
 TW Hya$^{a}$   & $(-16,\,+5)$   & $<$ 0.014& & 0.3 & 40 & 1000 &$<$ 0.00003 &$<$ 0.001 &&&&\\ 

 $\epsilon$ Cha\,{\sc I\,}N & $(+42,\,-24)$ & $<$ 1.071 &  &39& 50 & 2 &$<$ 10&$<$ 3&&&&\\

 Sa136 & $(-46,\,+146)$ & 1.62 (0.06) & $-$11 $\rightarrow$ 10  & 130 & 40 &6 & \hspace{0.28cm}13& \hspace{0.28cm}1&&100& \hspace{0.28cm}0.8& \hspace{0.28cm}0.06\\ 
       & $(-16,\,-4)$  & 4.03 (0.10) &$-$52 $\rightarrow$ 40  & 38 &40 &2& \hspace{0.28cm}74 &\hspace{0.28cm}20&&100& \hspace{0.28cm}1& \hspace{0.28cm}0.3\\ 
       & $(+13,\,-154)$ & 1.35 (0.05) & $-$16 $\rightarrow$ 3                  &280  & 40 &13& \hspace{0.28cm}4 &\hspace{0.28cm}0.1&&100& \hspace{0.28cm}0.5& \hspace{0.28cm}0.02\\ 

 HH54 B & $(+2,\,+5)$ & 0.91 (0.06) & $-$15 $\rightarrow$ 5  & 3 & 330 &200&\hspace{0.28cm}0.09 & \hspace{0.28cm}0.3&&&&\\ 

 G327.3-0.6 & $(-1,\,-21)$ & $<$ 1.56 &  & 2360 & 100 & 4 & $<$ 0.02 & \hspace{0.08cm}$<$ 0.00008&&&&\\ 
 NGC6334\,{\sc I}, \\ 
Case 1     & $(+13,\,-33)$    & 2.23 (0.15) & \hspace{0.10cm}$-$13 $\rightarrow$ $-$5  & 10 &100& 4& \hspace{0.28cm}5& \hspace{0.38cm}5&&&& \\
 NGC6334\,{\sc I},\\
 Case 2     & $(+13,\,-33)$    &96.0 (0.6) & \hspace{0.14cm}$-$93 $\rightarrow$ 36 & 10   &100& 4& \hspace{0.28cm}235 & \hspace{0.28cm}200&&&& \\
 
Ser SMM1&$(-3,\,+8)$& 0.36 (0.05) & $-$1 $\rightarrow$ 8  & 50 &35& 1& \hspace{0.28cm}48 & \hspace{0.28cm}9&&&& \\ 
       &      & 0.43 (0.05) & 8 $\rightarrow$ 17   & 50 &35& 1& \hspace{0.28cm}23 & \hspace{0.28cm}5&&&& \\ 

3C391 BML$^b$  & $(+1,\,-7)$  & $<$ 0.82& &40&50&10&$<$ 2& $<$ 0.4&&&&\\ 

 B335        & $(+3,\,-36)$     & $<$ 1.62      &     &  27 & 20 & 1& $<$ 337 & $<$ 100&&&& \\ 
 L1157           & $(-21,\,+121)$   & 0.08 (0.02) &$-$2 $\rightarrow$ 3  & 16 & 30 & 2& \hspace{0.28cm}5 & \hspace{0.28cm}3&&&& \\ 
                 &             & 1.75 (0.05) & 3 $\rightarrow$ 36  & 16 & 30 & 2& \hspace{0.28cm}88 & \hspace{0.28cm}60 &&&&\\ 
                 & $(-4,\,+64)$     & 0.71 (0.06) &$-$16 $\rightarrow$ 2  & 16 & 30 & 2& \hspace{0.28cm}38 & \hspace{0.28cm}20 &&&&\\ 
                 &             & 2.33 (0.08) & 2 $\rightarrow$ 38  & 16 & 30 & 2& \hspace{0.28cm}180 & \hspace{0.28cm}100 &&&&\\ 
                 & $(+12,\,+6)$      & 1.03 (0.04) &$-$15 $\rightarrow$ 3 & 20 & 30 & 3& \hspace{0.28cm}34 & \hspace{0.28cm}20&&1000&0.08&0.04 \\ 
                 &             & 0.21 (0.02) & 3 $\rightarrow$ 8   & 20 & 30 & 3& \hspace{0.28cm}7 & \hspace{0.28cm}3&&1000&0.02& 0.008\\ 
                 & $(+29,\,-52)$   & 1.83 (0.04) & $-$21 $\rightarrow$ 3 & 23 & 30 & 3& \hspace{0.28cm}53 & \hspace{0.28cm}20 &&500&0.3&0.1\\ 
                 &             & 0.29 (0.02) & 3 $\rightarrow$ 8   & 23 & 30 & 3& \hspace{0.28cm}9 & \hspace{0.28cm}4&&500&0.05& 0.02\\ 

    NGC7538 IRS 1       & $(+26,\,-4)$   & 2.11 (0.33) & $-$62 $\rightarrow$ $-$50 & 39 & 40 & 3& \hspace{0.28cm}51& \hspace{0.28cm}10&&&&\\ 
\noalign{\smallskip}
\noalign{\smallskip}
\hline                                             
\end{longtable}
Notes to the Table: $^{a}$ The volume density and column density are
inferred from HCN and \htva\ observations. $^{b}$ The volume density
and column density are inferred from OH 1720 MHz maser observations.
\end{landscape}
}

\begin{table*}[ht]
\begin{flushleft}
  \caption{Column densities of \ohtvao\ and estimates of the \orthowater\ abundance, $X$(\ohtvao) = $N$(\ohtvao)/$N$(\htva).}
%\resizebox{\hsize}{!}{
\begin{tabular}{l l c c c c r c c c}
\hline
\hline
\noalign{\smallskip}
 Source & Offset & $\int T_{\mathrm{mb}}$(H${}_2$O)$d \upsilon$ &  $\Delta \upsilon$  & $N\mathrm{(H_2)}$ & $N$(\ohtvao)& $X$(\ohtvao) \\
      & (\asec,\asec) & (K km s${}^{-1}$) &   & (\expo{19} cm${}^{-2}$) &  (\expo{15} $\mathrm{cm^{-2})}$ &(\expo{-7})\\                                
\noalign{\smallskip}
\hline
  \noalign{\smallskip}
 \noalign{\smallskip}
 \multicolumn{3}{l}{\textbf{Odin sources analyzed by other authors:}} \\
\noalign{\smallskip}
 W3$^{a}$       & $(0,\,0)$  & 3.9 (0.1)      &                            &&& \hspace{0.28cm}0.02 \\  
 Orion KL$^{b}$ & $(0,\,0)$     & 323 &  &   & 400  & 80 / 1000 \\  
               & $(+0,\,-240)$  & 24 &  &   & 0.09  & \hspace{0.28cm}0.1        \\ 
 Cha-MMS1$^{c}$ & $(+18,\,-21)$ &  &  &  &   & $<$ 0.07 \\           
                & $(-35,\,+32)$ &  &  &  &  & $<$ 0.07 \\ 
 IRAS 16293-2422$^{d}$ & $(+68,\,-50)$    &         \centering{}               & red wing                      && & \hspace{0.28cm}30 \\        
                       &             &                        & $\ \, $blue wing                    && & \hspace{0.05cm}11\\ 
                       & $(+31,\,-10)$    &                        & red wing                     && & \hspace{0.28cm}600 \\
                       &             &                     & $\ \, $blue wing                    && & \hspace{0.28cm}60 \\
                       & $(-23,\,-8)$    &                        & red wing                       && & 130 \\
                       &             &                     & $\ \, $blue wing                 && & 11\\
 S140 $^{e}$            &   $(0,\,0)$     &           &                              & && 0.05 -- 4 \\ 
 VLA1623$^{f}$     &     $(0,\,0)$        &                             & red wing & 2 & 0.2 -- 0.4 &100 -- 200 \\
  \noalign{\smallskip}
 \noalign{\smallskip}
  \hline
  \end{tabular}
%  }
\end{flushleft}
   Notes to the Table: $^{a}$ \citet{Wilson:2003fj}. The offset is with
   respect to \atwozero\ = 02:25:40.6, \dtwozero\ = +62:05:57. $^{b}$
   \citet{Olofsson:2003yq}. These authors use two different values for
   $N$(\htva) towards the (0, 0) position. The offsets are with respect
   to \atwozero\ = 05:35:14.4, \dtwozero\ = -05:22:30. $^{c}$
   \citet{Klotz:2008lr}. The offsets are with respect to \atwozero\ =
   11:06:31.7, \dtwozero\ = -77:23:32. $^{d}$
   \citet{Ristorcelli:2005fk}; Ristorcelli, private communication. The
   offsets are with respect to \atwozero\ = 16:32:23.0, \dtwozero\ =
   -24:28:40. $^{e}$ \citet{Persson:2009lr}, towards the PDR and the
   outflow. The offset is with respect to \atwozero\ = 22:19:19.4,
   \dtwozero\ = +63:18:50. $^{f}$ Larsson et al. in prep, private
   communication. The offset is with respect to \atwozero\ =
   16:23:20.2, \dtwozero\ = -24:22:55.
\end{table*}

\bibliographystyle{aa} % style aa.bst
\bibliography{/Users/per/Documents/LATEX/References/papers}

\begin{thebibliography}{86}
\expandafter\ifx\csname natexlab\endcsname\relax\def\natexlab#1{#1}\fi

\bibitem[{{Ashby} {et~al.}(2000){Ashby}, {Bergin}, {Plume}, {Carpenter},
  {Melnick}, {Chin}, {Erickson}, {Goldsmith}, {Harwit}, {Howe}, {Kleiner},
  {Koch}, {Neufeld}, {Patten}, {Schieder}, {Snell}, {Stauffer}, {Tolls},
  {Wang}, {Winnewisser}, \& {Zhang}}]{Ashby:2000fk}
{Ashby}, M.~L.~N., {Bergin}, E.~A., {Plume}, R., {et~al.} 2000, \apjl, 539,
  L119

\bibitem[{{Bachiller} {et~al.}(1995){Bachiller}, {Guilloteau}, {Dutrey},
  {Planesas}, \& {Martin-Pintado}}]{Bachiller:1995uq}
{Bachiller}, R., {Guilloteau}, S., {Dutrey}, A., {Planesas}, P., \&
  {Martin-Pintado}, J. 1995, \aap, 299, 857

\bibitem[{{Bachiller} {et~al.}(1987){Bachiller}, {Guilloteau}, \&
  {Kahane}}]{Bachiller:1987lr}
{Bachiller}, R., {Guilloteau}, S., \& {Kahane}, C. 1987, \aap, 173, 324

\bibitem[{{Bachiller} {et~al.}(1990){Bachiller}, {Martin-Pintado}, {Tafalla},
  {Cernicharo}, \& {Lazareff}}]{Bachiller:1990lr}
{Bachiller}, R., {Martin-Pintado}, J., {Tafalla}, M., {Cernicharo}, J., \&
  {Lazareff}, B. 1990, \aap, 231, 174

\bibitem[{{Bachiller} {et~al.}(2001){Bachiller}, {P{\'e}rez Guti{\'e}rrez},
  {Kumar}, \& {Tafalla}}]{Bachiller:2001lr}
{Bachiller}, R., {P{\'e}rez Guti{\'e}rrez}, M., {Kumar}, M.~S.~N., \&
  {Tafalla}, M. 2001, \aap, 372, 899

\bibitem[{{Benedettini} {et~al.}(2002){Benedettini}, {Viti}, {Giannini},
  {Nisini}, {Goldsmith}, \& {Saraceno}}]{Benedettini:2002fk}
{Benedettini}, M., {Viti}, S., {Giannini}, T., {et~al.} 2002, \aap, 395, 657

\bibitem[{{Bergin} {et~al.}(1998){Bergin}, {Neufeld}, \&
  {Melnick}}]{Bergin:1998lr}
{Bergin}, E.~A., {Neufeld}, D.~A., \& {Melnick}, G.~J. 1998, \apj, 499, 777

\bibitem[{{Bergman}(1992)}]{Bergman:1992kx}
{Bergman}, P. 1992, PhD thesis, G{\"o}teborg, Sweden, (1992)

\bibitem[{{Bourke}(2001)}]{Bourke:2001kx}
{Bourke}, T.~L. 2001, \apjl, 554, L91

\bibitem[{{Bourke} {et~al.}(1997){Bourke}, {Garay}, {Lehtinen}, {Koehnenkamp},
  {Launhardt}, {Nyman}, {May}, {Robinson}, \& {Hyland}}]{Bourke:1997uq}
{Bourke}, T.~L., {Garay}, G., {Lehtinen}, K.~K., {et~al.} 1997, \apj, 476, 781

\bibitem[{{Ceccarelli} {et~al.}(1997){Ceccarelli}, {Haas}, {Hollenbach}, \&
  {Rudolph}}]{Ceccarelli:1997kx}
{Ceccarelli}, C., {Haas}, M.~R., {Hollenbach}, D.~J., \& {Rudolph}, A.~L. 1997,
  \apj, 476, 771

\bibitem[{{Chen} \& {Slane}(2001)}]{Chen:2001fk}
{Chen}, Y. \& {Slane}, P.~O. 2001, \apj, 563, 202

\bibitem[{{Davis} {et~al.}(1999){Davis}, {Matthews}, {Ray}, {Dent}, \&
  {Richer}}]{Davis:1999fj}
{Davis}, C.~J., {Matthews}, H.~E., {Ray}, T.~P., {Dent}, W.~R.~F., \& {Richer},
  J.~S. 1999, \mnras, 309, 141

\bibitem[{{de la Reza} {et~al.}(2006){de la Reza}, {Jilinski}, \&
  {Ortega}}]{de-la-Reza:2006uq}
{de la Reza}, R., {Jilinski}, E., \& {Ortega}, V.~G. 2006, \aj, 131, 2609

\bibitem[{{de Lara} {et~al.}(1991){de Lara}, {Chavarria-K.}, \&
  {Lopez-Molina}}]{de-Lara:1991qy}
{de Lara}, E., {Chavarria-K.}, C., \& {Lopez-Molina}, G. 1991, \aap, 243, 139

\bibitem[{{Dickman}(1978)}]{Dickman:1978fk}
{Dickman}, R.~L. 1978, \apjs, 37, 407

\bibitem[{{Dubernet} {et~al.}(2009){Dubernet}, {Daniel}, {Grosjean}, \&
  {Lin}}]{Dubernet:2009lr}
{Dubernet}, M.-L., {Daniel}, F., {Grosjean}, A., \& {Lin}, C.~Y. 2009, \aap,
  497, 911

\bibitem[{{Dubernet} \& {Grosjean}(2002)}]{Dubernet:2002lr}
{Dubernet}, M.-L. \& {Grosjean}, A. 2002, \aap, 390, 793

\bibitem[{{Dupree} {et~al.}(2005){Dupree}, {Brickhouse}, {Smith}, \&
  {Strader}}]{Dupree:2005fk}
{Dupree}, A.~K., {Brickhouse}, N.~S., {Smith}, G.~H., \& {Strader}, J. 2005,
  \apjl, 625, L131

\bibitem[{{Enoch} {et~al.}(2006){Enoch}, {Young}, {Glenn}, {Evans}, {Golwala},
  {Sargent}, {Harvey}, {Aguirre}, {Goldin}, {Haig}, {Huard}, {Lange},
  {Laurent}, {Maloney}, {Mauskopf}, {Rossinot}, \& {Sayers}}]{Enoch:2006lr}
{Enoch}, M.~L., {Young}, K.~E., {Glenn}, J., {et~al.} 2006, \apj, 638, 293

\bibitem[{{Evans} {et~al.}(2005){Evans}, {Lee}, {Rawlings}, \&
  {Choi}}]{Evans:2005qy}
{Evans}, II, N.~J., {Lee}, J.-E., {Rawlings}, J.~M.~C., \& {Choi}, M. 2005,
  \apj, 626, 919

\bibitem[{{Faure} {et~al.}(2007){Faure}, {Crimier}, {Ceccarelli}, {Valiron},
  {Wiesenfeld}, \& {Dubernet}}]{Faure:2007lr}
{Faure}, A., {Crimier}, N., {Ceccarelli}, C., {et~al.} 2007, \aap, 472, 1029

\bibitem[{{Fesen}(1984)}]{Fesen:1984vn}
{Fesen}, R.~A. 1984, \apj, 281, 658

\bibitem[{{Franklin} {et~al.}(2008){Franklin}, {Snell}, {Kaufman}, {Melnick},
  {Neufeld}, {Hollenbach}, \& {Bergin}}]{Franklin:2008fk}
{Franklin}, J., {Snell}, R.~L., {Kaufman}, M.~J., {et~al.} 2008, \apj, 674,
  1015

\bibitem[{{Frisk} {et~al.}(2003){Frisk}, {Hagstr{\"o}m}, {Ala-Laurinaho},
  {Andersson}, {Berges}, {Chabaud}, {Dahlgren}, {Emrich}, {Flor{\'e}n},
  {Florin}, {Fredrixon}, {Gaier}, {Haas}, {Hirvonen}, {Hjalmarsson},
  {Jakobsson}, {Jukkala}, {Kildal}, {Kollberg}, {Lassing}, {Lecacheux},
  {Lehikoinen}, {Lehto}, {Mallat}, {Marty}, {Michet}, {Narbonne}, {Nexon},
  {Olberg}, {Olofsson}, {Olofsson}, {Orign{\'e}}, {Petersson}, {Piironen},
  {Pons}, {Pouliquen}, {Ristorcelli}, {Rosolen}, {Rouaix}, {R{\"a}is{\"a}nen},
  {Serra}, {Sj{\"o}berg}, {Stenmark}, {Torchinsky}, {Tuovinen}, {Ullberg},
  {Vinterhav}, {Wadefalk}, {Zirath}, {Zimmermann}, \&
  {Zimmermann}}]{Frisk:2003lr}
{Frisk}, U., {Hagstr{\"o}m}, M., {Ala-Laurinaho}, J., {et~al.} 2003, \aap, 402,
  L27

\bibitem[{{G{\aa}lfalk} \& {Olofsson}(2007)}]{Galfalk:2007lr}
{G{\aa}lfalk}, M. \& {Olofsson}, G. 2007, \aap, 475, 281

\bibitem[{{Giannini} {et~al.}(2006){Giannini}, {McCoey}, {Nisini}, {Cabrit},
  {Caratti o Garatti}, {Calzoletti}, \& {Flower}}]{Giannini:2006lr}
{Giannini}, T., {McCoey}, C., {Nisini}, B., {et~al.} 2006, \aap, 459, 821

\bibitem[{{Gueth} \& {Guilloteau}(1999)}]{Gueth:1999vn}
{Gueth}, F. \& {Guilloteau}, S. 1999, \aap, 343, 571

\bibitem[{{Gusdorf} {et~al.}(2008){Gusdorf}, {Cabrit}, {Flower}, \& {Pineau Des
  For{\^e}ts}}]{Gusdorf:2008lr}
{Gusdorf}, A., {Cabrit}, S., {Flower}, D.~R., \& {Pineau Des For{\^e}ts}, G.
  2008, \aap, 482, 809

\bibitem[{{Hartstein} \& {Liseau}(1998)}]{Hartstein:1998fk}
{Hartstein}, D. \& {Liseau}, R. 1998, \aap, 332, 703

\bibitem[{{Henning} {et~al.}(1993){Henning}, {Pfau}, {Zinnecker}, \&
  {Prusti}}]{Henning:1993lr}
{Henning}, T., {Pfau}, W., {Zinnecker}, H., \& {Prusti}, T. 1993, \aap, 276,
  129

\bibitem[{{Herbig}(1998)}]{Herbig:1998fk}
{Herbig}, G.~H. 1998, \apj, 497, 736

\bibitem[{{Herczeg} {et~al.}(2004){Herczeg}, {Wood}, {Linsky}, {Valenti}, \&
  {Johns-Krull}}]{Herczeg:2004fj}
{Herczeg}, G.~J., {Wood}, B.~E., {Linsky}, J.~L., {Valenti}, J.~A., \&
  {Johns-Krull}, C.~M. 2004, \apj, 607, 369

\bibitem[{{Hirano} {et~al.}(1988){Hirano}, {Kameya}, {Nakayama}, \&
  {Takakubo}}]{Hirano:1988lr}
{Hirano}, N., {Kameya}, O., {Nakayama}, M., \& {Takakubo}, K. 1988, \apjl, 327,
  L69

\bibitem[{{Hjalmarson} {et~al.}(2003){Hjalmarson}, {Frisk}, {Olberg},
  {Bergman}, {Bernath}, {Biver}, {Black}, {Booth}, {Buat}, {Crovisier},
  {Curry}, {Dahlgren}, {Encrenaz}, {Falgarone}, {Feldman}, {Fich},
  {Flor{\'e}n}, {Fredrixon}, {Gerin}, {Gregersen}, {Hagstr{\"o}m}, {Harju},
  {Hasegawa}, {Horellou}, {Johansson}, {Kyr{\"o}l{\"a}}, {Kwok}, {Larsson},
  {Lecacheux}, {Liljestr{\"o}m}, {Lindqvist}, {Liseau}, {Llewellyn}, {Mattila},
  {M{\'e}gie}, {Mitchell}, {Murtagh}, {Nyman}, {Nordh}, {Olofsson}, {Olofsson},
  {Olofsson}, {Pagani}, {Persson}, {Plume}, {Rickman}, {Ristorcelli},
  {Rydbeck}, {Sandqvist}, {von Sch{\'e}ele}, {Serra}, {Torchinsky}, {Tothill},
  {Volk}, {Wiklind}, {Wilson}, {Winnberg}, \& {Witt}}]{Hjalmarson:2003kx}
{Hjalmarson}, {\AA}., {Frisk}, U., {Olberg}, M., {et~al.} 2003, \aap, 402, L39

\bibitem[{{Hughes} \& {Hartigan}(1992)}]{Hughes:1992qy}
{Hughes}, J. \& {Hartigan}, P. 1992, \aj, 104, 680

\bibitem[{{Kameya} {et~al.}(1989){Kameya}, {Hasegawa}, {Hirano}, {Takakubo}, \&
  {Seki}}]{Kameya:1989fj}
{Kameya}, O., {Hasegawa}, T.~I., {Hirano}, N., {Takakubo}, K., \& {Seki}, M.
  1989, \apj, 339, 222

\bibitem[{{Kastner} {et~al.}(1997){Kastner}, {Zuckerman}, {Weintraub}, \&
  {Forveille}}]{Kastner:1997kx}
{Kastner}, J.~H., {Zuckerman}, B., {Weintraub}, D.~A., \& {Forveille}, T. 1997,
  Science, 277, 67

\bibitem[{{Kaufman} \& {Neufeld}(1996)}]{Kaufman:1996qy}
{Kaufman}, M.~J. \& {Neufeld}, D.~A. 1996, \apj, 456, 611

\bibitem[{{Kenyon} {et~al.}(1994){Kenyon}, {Dobrzycka}, \&
  {Hartmann}}]{Kenyon:1994uq}
{Kenyon}, S.~J., {Dobrzycka}, D., \& {Hartmann}, L. 1994, \aj, 108, 1872

\bibitem[{{Kessler} {et~al.}(1996){Kessler}, {Steinz}, {Anderegg}, {Clavel},
  {Drechsel}, {Estaria}, {Faelker}, {Riedinger}, {Robson}, {Taylor}, \&
  {Xim{\'e}nez de Ferr{\'a}n}}]{Kessler:1996fj}
{Kessler}, M.~F., {Steinz}, J.~A., {Anderegg}, M.~E., {et~al.} 1996, \aap, 315,
  L27

\bibitem[{{Klotz} {et~al.}(2008){Klotz}, {Harju}, {Ristorcelli}, {Juvela},
  {Boudet}, \& {Haikala}}]{Klotz:2008lr}
{Klotz}, A., {Harju}, J., {Ristorcelli}, I., {et~al.} 2008, \aap, 488, 559

\bibitem[{{Knee}(1992)}]{Knee:1992lr}
{Knee}, L.~B.~G. 1992, \aap, 259, 283

\bibitem[{{Knude} \& {Hog}(1998)}]{Knude:1998qy}
{Knude}, J. \& {Hog}, E. 1998, \aap, 338, 897

\bibitem[{{Lamzin} {et~al.}(2004){Lamzin}, {Kravtsova}, {Romanova}, \&
  {Batalha}}]{Lamzin:2004qy}
{Lamzin}, S.~A., {Kravtsova}, A.~S., {Romanova}, M.~M., \& {Batalha}, C. 2004,
  Astronomy Letters, 30, 413

\bibitem[{{Larsson} {et~al.}(2002){Larsson}, {Liseau}, \&
  {Men'shchikov}}]{Larsson:2002lr}
{Larsson}, B., {Liseau}, R., \& {Men'shchikov}, A.~B. 2002, \aap, 386, 1055

\bibitem[{{Lee} {et~al.}(2007){Lee}, {Ho}, {Palau}, {Hirano}, {Bourke},
  {Shang}, \& {Zhang}}]{Lee:2007kx}
{Lee}, C.-F., {Ho}, P.~T.~P., {Palau}, A., {et~al.} 2007, \apj, 670, 1188

\bibitem[{{Leurini} {et~al.}(2006){Leurini}, {Schilke}, {Parise}, {Wyrowski},
  {G{\"u}sten}, \& {Philipp}}]{Leurini:2006uq}
{Leurini}, S., {Schilke}, P., {Parise}, B., {et~al.} 2006, \aap, 454, L83

\bibitem[{{Linke} {et~al.}(1977){Linke}, {Goldsmith}, {Wannier}, {Wilson}, \&
  {Penzias}}]{Linke:1977lr}
{Linke}, R.~A., {Goldsmith}, P.~F., {Wannier}, P.~G., {Wilson}, R.~W., \&
  {Penzias}, A.~A. 1977, \apj, 214, 50

\bibitem[{{Liseau} {et~al.}(1996){Liseau}, {Ceccarelli}, {Larsson}, {Nisini},
  {White}, {Ade}, {Armand}, {Burgdorf}, {Caux}, {Cerulli}, {Church}, {Clegg},
  {Digorgio}, {Furniss}, {Giannini}, {Glencross}, {Gry}, {King}, {Lim},
  {Lorenzetti}, {Molinari}, {Naylor}, {Orfei}, {Saraceno}, {Sidher}, {Smith},
  {Spinoglio}, {Swinyard}, {Texier}, {Tommasi}, {Trams}, \&
  {Unger}}]{Liseau:1996fk}
{Liseau}, R., {Ceccarelli}, C., {Larsson}, B., {et~al.} 1996, \aap, 315, L181

\bibitem[{{Liseau} {et~al.}(2005){Liseau}, {Fridlund}, \&
  {Larsson}}]{Liseau:2005lr}
{Liseau}, R., {Fridlund}, C.~V.~M., \& {Larsson}, B. 2005, \apj, 619, 959

\bibitem[{{Liseau} \& {Olofsson}(1999)}]{Liseau:1999uq}
{Liseau}, R. \& {Olofsson}, G. 1999, \aap, 343, L83

\bibitem[{{Lockett} {et~al.}(1999){Lockett}, {Gauthier}, \&
  {Elitzur}}]{Lockett:1999qy}
{Lockett}, P., {Gauthier}, E., \& {Elitzur}, M. 1999, \apj, 511, 235

\bibitem[{{Looney} {et~al.}(2007){Looney}, {Tobin}, \& {Kwon}}]{Looney:2007lr}
{Looney}, L.~W., {Tobin}, J.~J., \& {Kwon}, W. 2007, \apjl, 670, L131

\bibitem[{{Mattila} {et~al.}(1989){Mattila}, {Liljestr{\"o}m}, \&
  {Toriseva}}]{Mattila:1989lr}
{Mattila}, K., {Liljestr{\"o}m}, T., \& {Toriseva}, M. 1989, in Low Mass Star
  Formation and Pre-main Sequence Objects, ed. B.~{Reipurth}, 153--171

\bibitem[{{McCutcheon} {et~al.}(2000){McCutcheon}, {Sandell}, {Matthews},
  {Kuiper}, {Sutton}, {Danchi}, \& {Sato}}]{McCutcheon:2000yq}
{McCutcheon}, W.~H., {Sandell}, G., {Matthews}, H.~E., {et~al.} 2000, \mnras,
  316, 152

\bibitem[{{Melnick} {et~al.}(2000){Melnick}, {Stauffer}, {Ashby}, {Bergin},
  {Chin}, {Erickson}, {Goldsmith}, {Harwit}, {Howe}, {Kleiner}, {Koch},
  {Neufeld}, {Patten}, {Plume}, {Schieder}, {Snell}, {Tolls}, {Wang},
  {Winnewisser}, \& {Zhang}}]{Melnick:2000fk}
{Melnick}, G.~J., {Stauffer}, J.~R., {Ashby}, M.~L.~N., {et~al.} 2000, \apjl,
  539, L77

\bibitem[{{Moscadelli} {et~al.}(2008){Moscadelli}, {Reid}, {Menten},
  {Brunthaler}, {Zheng}, \& {Xu}}]{Moscadelli:2008uq}
{Moscadelli}, L., {Reid}, M.~J., {Menten}, K.~M., {et~al.} 2008, ArXiv e-prints

\bibitem[{{Neckel}(1978)}]{Neckel:1978fj}
{Neckel}, T. 1978, \aap, 69, 51

\bibitem[{{Neufeld} {et~al.}(2006){Neufeld}, {Melnick}, {Sonnentrucker},
  {Bergin}, {Green}, {Kim}, {Watson}, {Forrest}, \& {Pipher}}]{Neufeld:2006fk}
{Neufeld}, D.~A., {Melnick}, G.~J., {Sonnentrucker}, P., {et~al.} 2006, \apj,
  649, 816

\bibitem[{{Nisini} {et~al.}(2007){Nisini}, {Codella}, {Giannini}, {Santiago
  Garcia}, {Richer}, {Bachiller}, \& {Tafalla}}]{Nisini:2007fk}
{Nisini}, B., {Codella}, C., {Giannini}, T., {et~al.} 2007, \aap, 462, 163

\bibitem[{{Nordh} {et~al.}(2003){Nordh}, {von Sch{\'e}ele}, {Frisk}, {Ahola},
  {Booth}, {Encrenaz}, {Hjalmarson}, {Kendall}, {Kyr{\"o}l{\"a}}, {Kwok},
  {Lecacheux}, {Leppelmeier}, {Llewellyn}, {Mattila}, {M{\'e}gie}, {Murtagh},
  {Rougeron}, \& {Witt}}]{Nordh:2003qy}
{Nordh}, H.~L., {von Sch{\'e}ele}, F., {Frisk}, U., {et~al.} 2003, \aap, 402,
  L21

\bibitem[{{Olberg} {et~al.}(2003){Olberg}, {Frisk}, {Lecacheux}, {Olofsson},
  {Baron}, {Bergman}, {Florin}, {Hjalmarson}, {Larsson}, {Murtagh}, {Olofsson},
  {Pagani}, {Sandqvist}, {Teyssier}, {Torchinsky}, \& {Volk}}]{Olberg:2003fj}
{Olberg}, M., {Frisk}, U., {Lecacheux}, A., {et~al.} 2003, \aap, 402, L35

\bibitem[{{Olofsson} {et~al.}(2003){Olofsson}, {Olofsson}, {Hjalmarson},
  {Bergman}, {Black}, {Booth}, {Buat}, {Curry}, {Encrenaz}, {Falgarone},
  {Feldman}, {Fich}, {Flor{\'e}n}, {Frisk}, {Gerin}, {Gregersen}, {Harju},
  {Hasegawa}, {Johansson}, {Kwok}, {Larsson}, {Lecacheux}, {Liljestr{\"o}m},
  {Liseau}, {Mattila}, {Mitchell}, {Nordh}, {Olberg}, {Olofsson}, {Pagani},
  {Plume}, {Ristorcelli}, {Rydbeck}, {Sandqvist}, {von Sch{\'e}ele}, {Serra},
  {Tothill}, {Volk}, \& {Wilson}}]{Olofsson:2003yq}
{Olofsson}, A.~O.~H., {Olofsson}, G., {Hjalmarson}, {\AA}., {et~al.} 2003,
  \aap, 402, L47

\bibitem[{{Parise} {et~al.}(2008){Parise}, {Belloche}, {Leurini}, \&
  {Schilke}}]{Parise:2008rt}
{Parise}, B., {Belloche}, A., {Leurini}, S., \& {Schilke}, P. 2008, \apss, 313,
  73

\bibitem[{{Parise} {et~al.}(2006){Parise}, {Belloche}, {Leurini}, {Schilke},
  {Wyrowski}, \& {G{\"u}sten}}]{Parise:2006fk}
{Parise}, B., {Belloche}, A., {Leurini}, S., {et~al.} 2006, \aap, 454, L79

\bibitem[{{Persson} {et~al.}(2009){Persson}, {Olberg}, {.~Hjalmarson},
  {Spaans}, {Black}, {Frisk}, {Liljestr{\"o}m}, {Olofsson}, {Poelman}, \&
  {Sandqvist}}]{Persson:2009lr}
{Persson}, C.~M., {Olberg}, M., {.~Hjalmarson}, {\~A}., {et~al.} 2009, \aap,
  494, 637

\bibitem[{{Phillips} {et~al.}(1996){Phillips}, {Maluendes}, \&
  {Green}}]{Phillips:1996qy}
{Phillips}, T.~R., {Maluendes}, S., \& {Green}, S. 1996, \apjs, 107, 467

\bibitem[{{Qi} {et~al.}(2008){Qi}, {Wilner}, {Aikawa}, {Blake}, \&
  {Hogerheijde}}]{Qi:2008lr}
{Qi}, C., {Wilner}, D.~J., {Aikawa}, Y., {Blake}, G.~A., \& {Hogerheijde},
  M.~R. 2008, \apj, 681, 1396

\bibitem[{{Ristorcelli} {et~al.}(2005){Ristorcelli}, {Falgarone},
  {Sch{\"o}ier}, {Cabrit}, {Gerin}, {Baron}, {Frisk}, {Harju}, {Hjalmarson},
  {Klotz}, {Larsson}, {Liseau}, \& {Pagani}}]{Ristorcelli:2005fk}
{Ristorcelli}, I., {Falgarone}, E., {Sch{\"o}ier}, F., {et~al.} 2005, in IAU
  Symposium, Vol. 235, IAU Symposium, 227P--+

\bibitem[{{Sandell}(2000)}]{Sandell:2000fk}
{Sandell}, G. 2000, \aap, 358, 242

\bibitem[{{Sandell} \& {Sievers}(2004)}]{Sandell:2004qy}
{Sandell}, G. \& {Sievers}, A. 2004, \apj, 600, 269

\bibitem[{{Sandqvist}(1977)}]{Sandqvist:1977lr}
{Sandqvist}, A. 1977, \aap, 57, 467

\bibitem[{{Snell} {et~al.}(2005){Snell}, {Hollenbach}, {Howe}, {Neufeld},
  {Kaufman}, {Melnick}, {Bergin}, \& {Wang}}]{Snell:2005lr}
{Snell}, R.~L., {Hollenbach}, D., {Howe}, J.~E., {et~al.} 2005, \apj, 620, 758

\bibitem[{{Snell} {et~al.}(2000){Snell}, {Howe}, {Ashby}, {Bergin}, {Chin},
  {Erickson}, {Goldsmith}, {Harwit}, {Kleiner}, {Koch}, {Neufeld}, {Patten},
  {Plume}, {Schieder}, {Stauffer}, {Tolls}, {Wang}, {Winnewisser}, {Zhang}, \&
  {Melnick}}]{Snell:2000lr}
{Snell}, R.~L., {Howe}, J.~E., {Ashby}, M.~L.~N., {et~al.} 2000, \apjl, 539,
  L93

\bibitem[{{Snell} {et~al.}(1980){Snell}, {Loren}, \& {Plambeck}}]{Snell:1980lr}
{Snell}, R.~L., {Loren}, R.~B., \& {Plambeck}, R.~L. 1980, \apjl, 239, L17

\bibitem[{{Stojimirovi{\'c}} {et~al.}(2006){Stojimirovi{\'c}}, {Narayanan},
  {Snell}, \& {Bally}}]{Stojimirovic:2006fk}
{Stojimirovi{\'c}}, I., {Narayanan}, G., {Snell}, R.~L., \& {Bally}, J. 2006,
  \apj, 649, 280

\bibitem[{{Stutz} {et~al.}(2008){Stutz}, {Rubin}, {Werner}, {Rieke}, {Bieging},
  {Keene}, {Kang}, {Shirley}, {Su}, {Velusamy}, \& {Wilner}}]{Stutz:2008fk}
{Stutz}, A.~M., {Rubin}, M., {Werner}, M.~W., {et~al.} 2008, \apj, 687, 389

\bibitem[{{Ulich} \& {Haas}(1976)}]{Ulich:1976fk}
{Ulich}, B.~L. \& {Haas}, R.~W. 1976, \apjs, 30, 247

\bibitem[{{van der Tak} {et~al.}(2007){van der Tak}, {Black}, {Sch{\"o}ier},
  {Jansen}, \& {van Dishoeck}}]{van-der-Tak:2007fk}
{van der Tak}, F.~F.~S., {Black}, J.~H., {Sch{\"o}ier}, F.~L., {Jansen}, D.~J.,
  \& {van Dishoeck}, E.~F. 2007, \aap, 468, 627

\bibitem[{{van Dishoeck} {et~al.}(1993){van Dishoeck}, {Jansen}, \&
  {Phillips}}]{van-Dishoeck:1993fk}
{van Dishoeck}, E.~F., {Jansen}, D.~J., \& {Phillips}, T.~G. 1993, \aap, 279,
  541

\bibitem[{{van Zadelhoff} {et~al.}(2001){van Zadelhoff}, {van Dishoeck}, {Thi},
  \& {Blake}}]{van-Zadelhoff:2001yq}
{van Zadelhoff}, G.-J., {van Dishoeck}, E.~F., {Thi}, W.-F., \& {Blake}, G.~A.
  2001, \aap, 377, 566

\bibitem[{{Werner} {et~al.}(2004){Werner}, {Roellig}, {Low}, {Rieke}, {Rieke},
  {Hoffmann}, {Young}, {Houck}, {Brandl}, {Fazio}, {Hora}, {Gehrz}, {Helou},
  {Soifer}, {Stauffer}, {Keene}, {Eisenhardt}, {Gallagher}, {Gautier}, {Irace},
  {Lawrence}, {Simmons}, {Van Cleve}, {Jura}, {Wright}, \&
  {Cruikshank}}]{Werner:2004fk}
{Werner}, M.~W., {Roellig}, T.~L., {Low}, F.~J., {et~al.} 2004, \apjs, 154, 1

\bibitem[{{White} {et~al.}(1995){White}, {Casali}, \& {Eiroa}}]{White:1995lr}
{White}, G.~J., {Casali}, M.~M., \& {Eiroa}, C. 1995, \aap, 298, 594

\bibitem[{{Wilson} {et~al.}(2003){Wilson}, {Mason}, {Gregersen}, {Olofsson},
  {Bergman}, {Booth}, {Boudet}, {Buat}, {Curry}, {Encrenaz}, {Falgarone},
  {Feldman}, {Fich}, {Floren}, {Frisk}, {Gerin}, {Harju}, {Hasegawa},
  {Hjalmarson}, {Juvela}, {Kwok}, {Larsson}, {Lecacheux}, {Liljestrom},
  {Liseau}, {Mattila}, {Mitchell}, {Nordh}, {Olberg}, {Olofsson}, {Pagani},
  {Plume}, {Ristorcelli}, {Sandqvist}, {Serra}, {Tothill}, {Volk}, \& {von
  Scheele}}]{Wilson:2003fj}
{Wilson}, C.~D., {Mason}, A., {Gregersen}, E., {et~al.} 2003, \aap, 402, L59

\bibitem[{{Wyrowski} {et~al.}(2006){Wyrowski}, {Menten}, {Schilke},
  {Thorwirth}, {G{\"u}sten}, \& {Bergman}}]{Wyrowski:2006fj}
{Wyrowski}, F., {Menten}, K.~M., {Schilke}, P., {et~al.} 2006, \aap, 454, L91

\end{thebibliography}
\newpage
\newpage
\newpage

\begin{appendix}
\section{Ground based observations}
\label{appendix:CO}
The observations with the Swedish ESO Submillimetre Telescope (\sest)
were made during 11-20 August 1997 and 2-6 August 1998. Some
complementary SiO\,(2-1) map data were collected during 7--9 February
2003. The observed molecules and their transitions are listed in
Table\,\ref{SEST_obs} and for SiO\,(2-1) and (3-2), the observations
were done simultaneously. 

SIS receivers were used as frontends and the
backend was a $2\times 1$\,GHz multi-channel acousto-optical
spectrometer (AOS), split into two halves for the SiO\,(2-1) and (3-2)
observations. For the SiO\,(5-4) and the CO observations, both the
high resolution (HRS) and the low resolution (LRS) backends were
used. The channel separation of the HRS was 43\,kHz and the spectral
resolution 80\,kHz covering the bandwidth of 86\,MHz, corresponding to
$\delta \upsilon = 0.07$\,\kmpers\ and $\Delta \upsilon = \pm
35$\,\kmpers\ for the velocity resolution and range at 345\,GHz,
respectively.  For the LRS, these parameters were channel separation
0.69\,MHz and $\Delta \nu = 1.4$\,MHz spectral resolution, providing
$\delta \upsilon = 1.2$\,\kmpers\ and $\Delta \upsilon = \pm
700$\,\kmpers, respectively, for the CO\,(3-2) observations.  

The data were chopper-wheel calibrated in the $T_{\rm
  A}^{\star}$-scale \citep{Ulich:1976fk} and the main beam
efficiencies at the different frequencies, $\eta_{\rm mb}$, are given
in Table~\ref{SEST_obs}. The pointing of the telescope was regularly
checked towards point sources, masing in the SiO\,(v=1, $J$=2-1) line,
and was determined to be better than 3\asec\ (rms). However, for
HH\,54, all SiO data refer to the vibrational ground state, v=0, and
the (2-1) and (3-2) data were obtained in frequency switching mode,
with a frequency chop of 4\,MHz.  

\citet{Knee:1992lr} assigned a kinetic gas temperature of $\sim$\,15 K
to the bulk cloud material. In the CO (3-2) line, this should yield a
high contrast between the cloud and the high velocity gas. To achieve
flat, optimum baselines, the CO (3-2) observations were done therefore
in wide dual beam switching with throws of $\pm 11$\amin\ in
azimuth. At this maximum amplitude available at the \sest, the
reference beams were still inside the molecular cloud. For this
reason, this mode could not be adopted for the CO (2-1) observations,
which were performed in total power mode. The reference position was
1\adeg\ north of HH\,54.  Centered on the object, nine point maps with
25\asec\ spacings were obtained in both CO lines. In addition, a
tighter sampled nine point map with 15\asec\ spacings was also made in
CO\,(3-2).

\begin{table*}[ht]
\begin{flushleft}
  \caption{\label{SEST_obs} Molecular line observations with the 15\,m
    SEST}
%\resizebox{\hsize}{!}{
\begin{tabular}{rccccclcl}
\hline
\hline
\noalign{\smallskip}
Transition & Frequency  &  HPBW   & $\eta_{\rm mb}$  & $T_{\rm sys}$ & $T_{\rm rms}$ & Backend  &  $\Delta \upsilon_{\rm lsr}$ & $\int\!\!T_{\rm A}^{\star}\,d\upsilon$ \\  
                  &   (MHz)         & (\asec)   &                                 & (K)    & (mK)  &                   &  (\kmpers)                                    & (K \kmpers)                                               \\ 
\noalign{\smallskip}
\hline
  \noalign{\smallskip}
 \noalign{\smallskip}
CO (2-1)  & 230538.0         &  23      & 0.50             &  \phantom{1}500    &   95/10   & HRS/LRS           & $-26.3$ to $-0.3$  & $15.6 \pm 0.3$     \\
CO (3-2)  & 345796.0         &  15      & 0.25             & 1600                        &  470/70  & HRS/LRS            & $-16.4$ to $+0.8$ & $13.6 \pm 1.4$      \\
SiO (2-1) & \phantom{1}86846.9 & 58 & 0.75       &  \phantom{1}160    & 10$^{a}$           & HRS          & &  \\
SiO (3-2) & 130268.7        &  38      & 0.68              &  \phantom{1}190    & 15$^{a}$          & HRS          & &  \\
SiO (5-4) & 217104.9        &  23      & 0.61              &  \phantom{1}750    & 35/7$^{a}$       & HRS/LRS & &  \\
  \noalign{\smallskip}
 \noalign{\smallskip}
  \hline
  \end{tabular}
%  }
\end{flushleft}
Note to the Table: $^{a}$ This is the rms of the average spectrum, including all positions.
\end{table*}
\end{appendix}

\Online

\begin{appendix}

\section{Online material}
For comparision we show the calibrated rawdata together with the
baseline subtracted and smoothed spectra already shown in the
text. 
\setcounter{figure}{0}
\begin{figure*}
\begin{flushleft}
 \includegraphics[width=7.5cm]{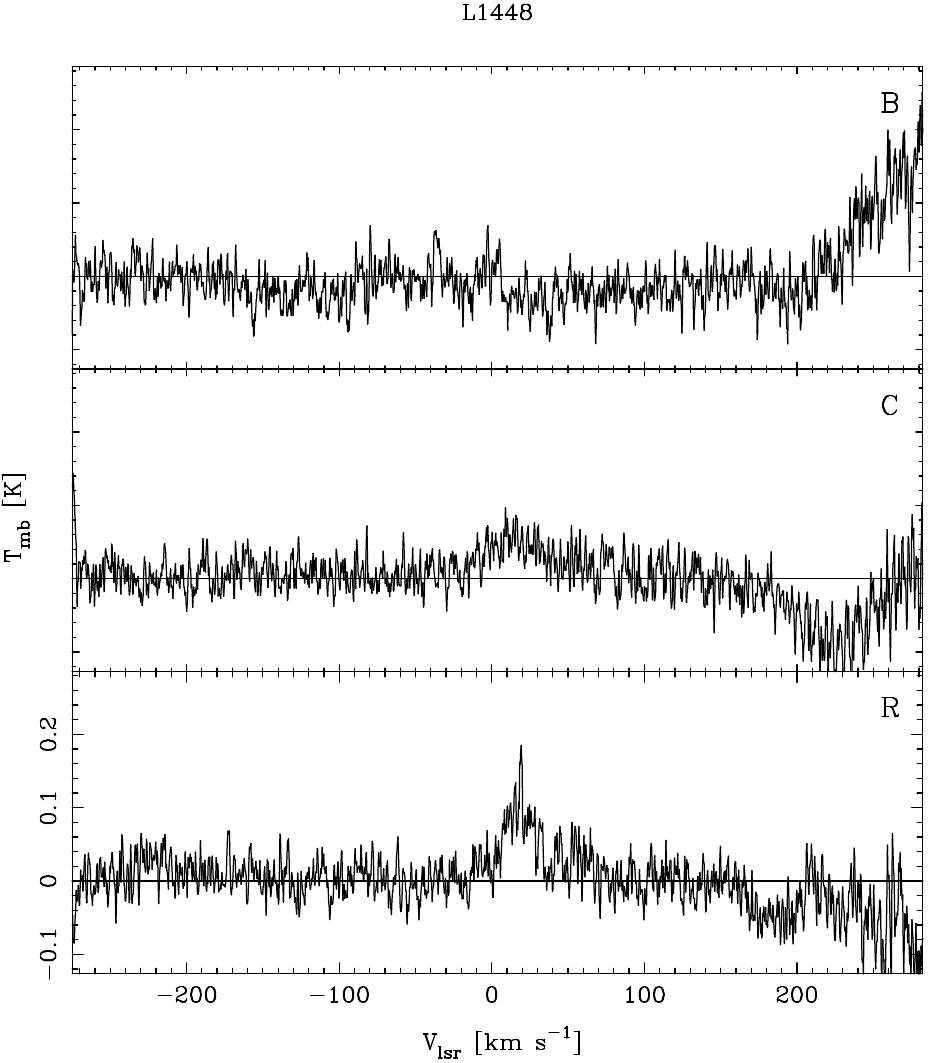} 
 \hspace{2cm}
 \includegraphics[width=7.5cm]{figureL1448.pdf}
   \\ \vspace{1cm}
   \includegraphics[width=7.5cm]{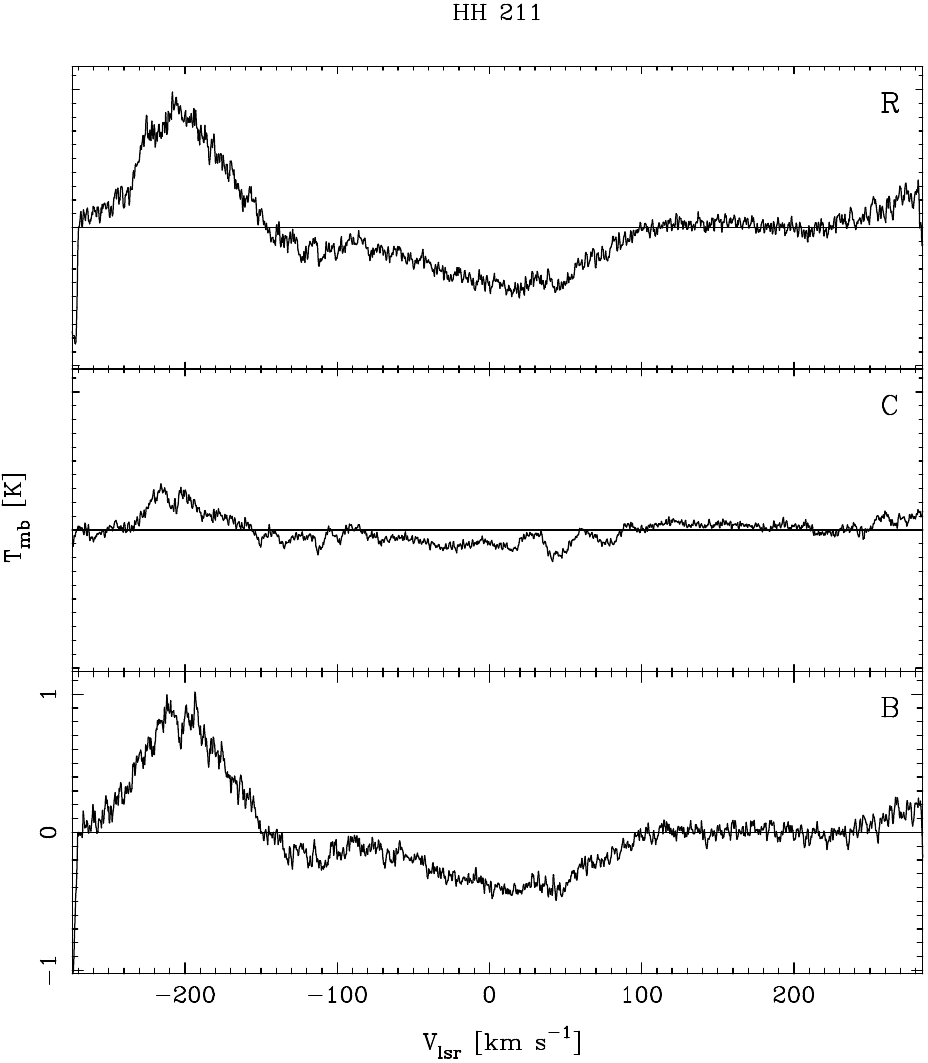} 
 \hspace{2cm}
 \includegraphics[width=7.5cm]{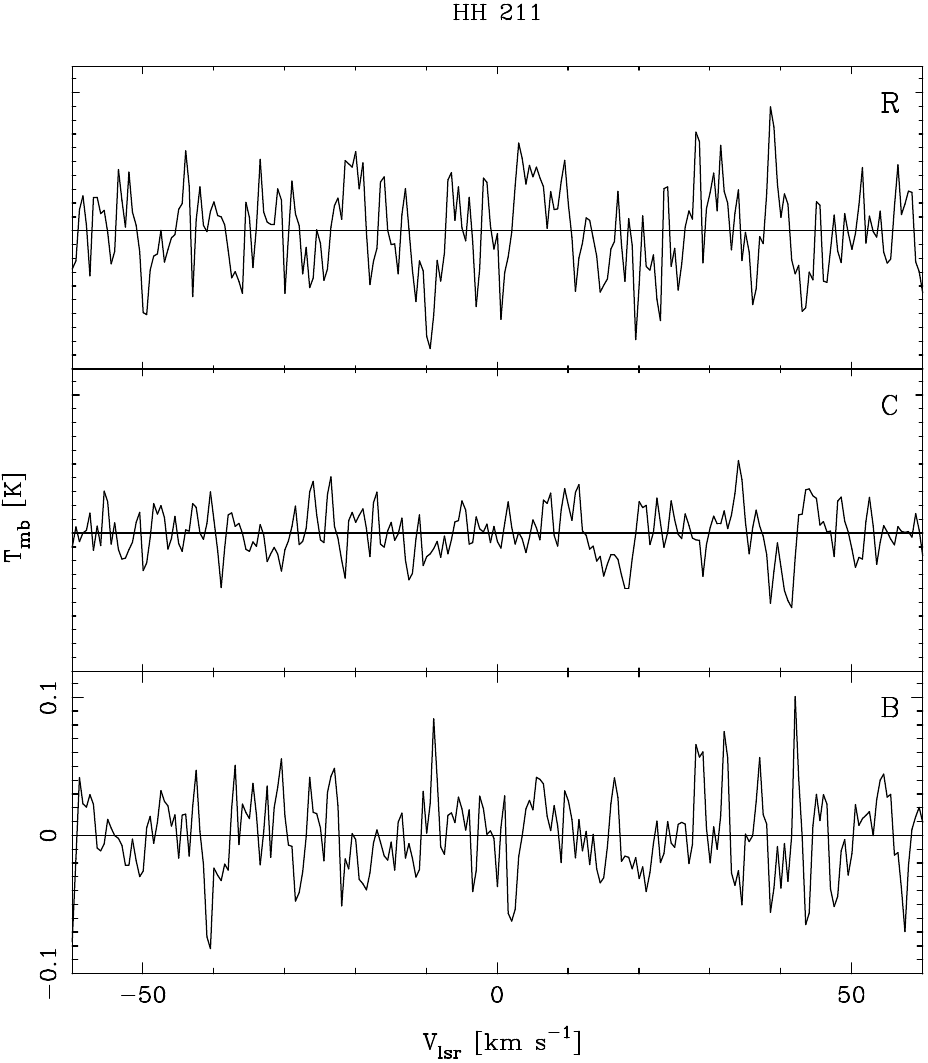}
 \\ \vspace{1cm}
   \includegraphics[width=7.5cm]{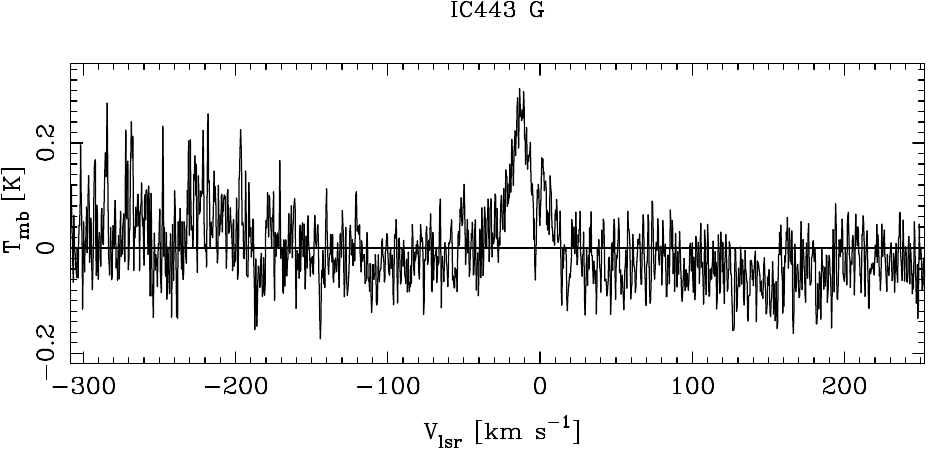} 
 \hspace{2cm}
 \includegraphics[width=7.5cm]{figureIC443-G.pdf}
 \caption{This figure shows the L1448, HH211 and IC443-G spectra. The
   positions are listed in Table~\ref{tab:table1}. All spectra on the
   right are smoothed to a resolution of 0.5 \kmpers\ and clipped for
   clarity. These spectra are also calibrated in frequency. The
   spectra on the left are the calibrated rawdata with a zero order
   baseline subtraction. For strip maps, a letter in the upper right
   corner indicates in which part of the flow the spectra has been
   collected (R=red, B=blue and C=center).}
     \label{fig:waterspectra}
\end{flushleft}
\end{figure*}

\begin{figure*}
\begin{flushleft}
 \includegraphics[width=7.5cm]{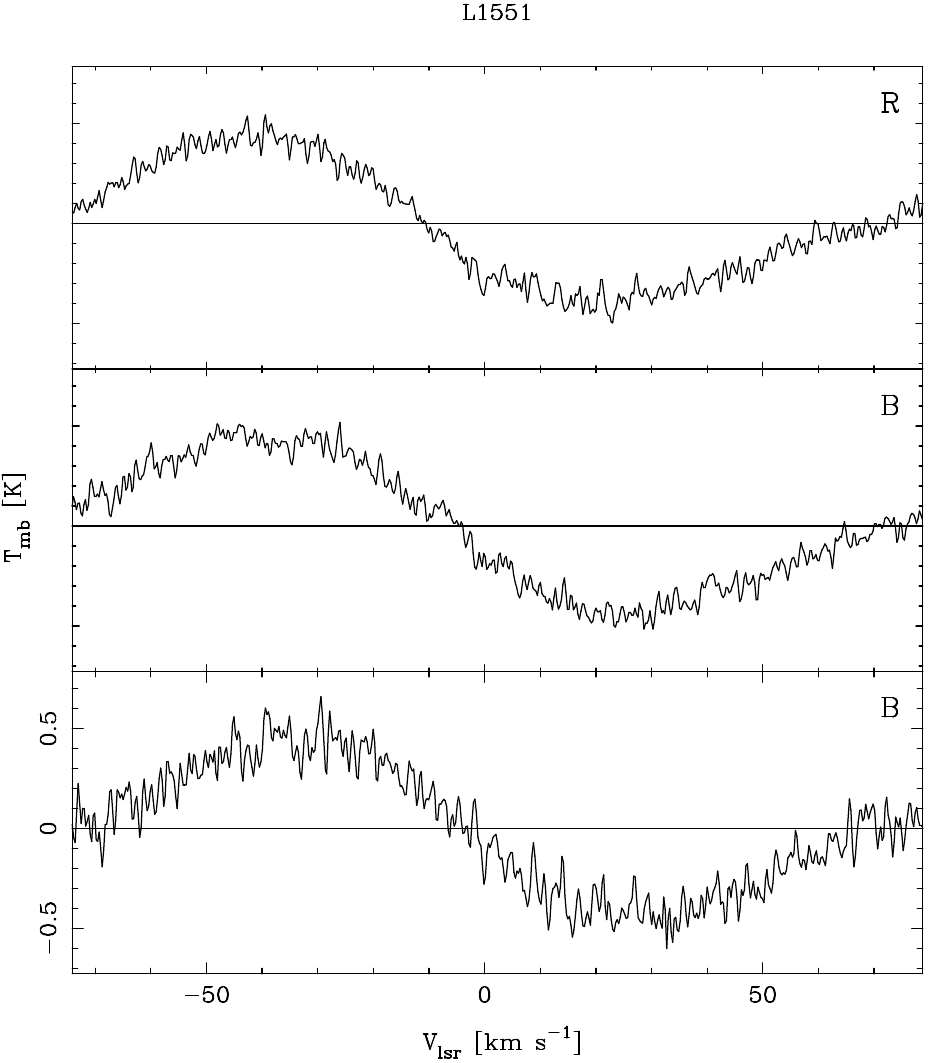} 
 \hspace{2cm}
 \includegraphics[width=7.5cm]{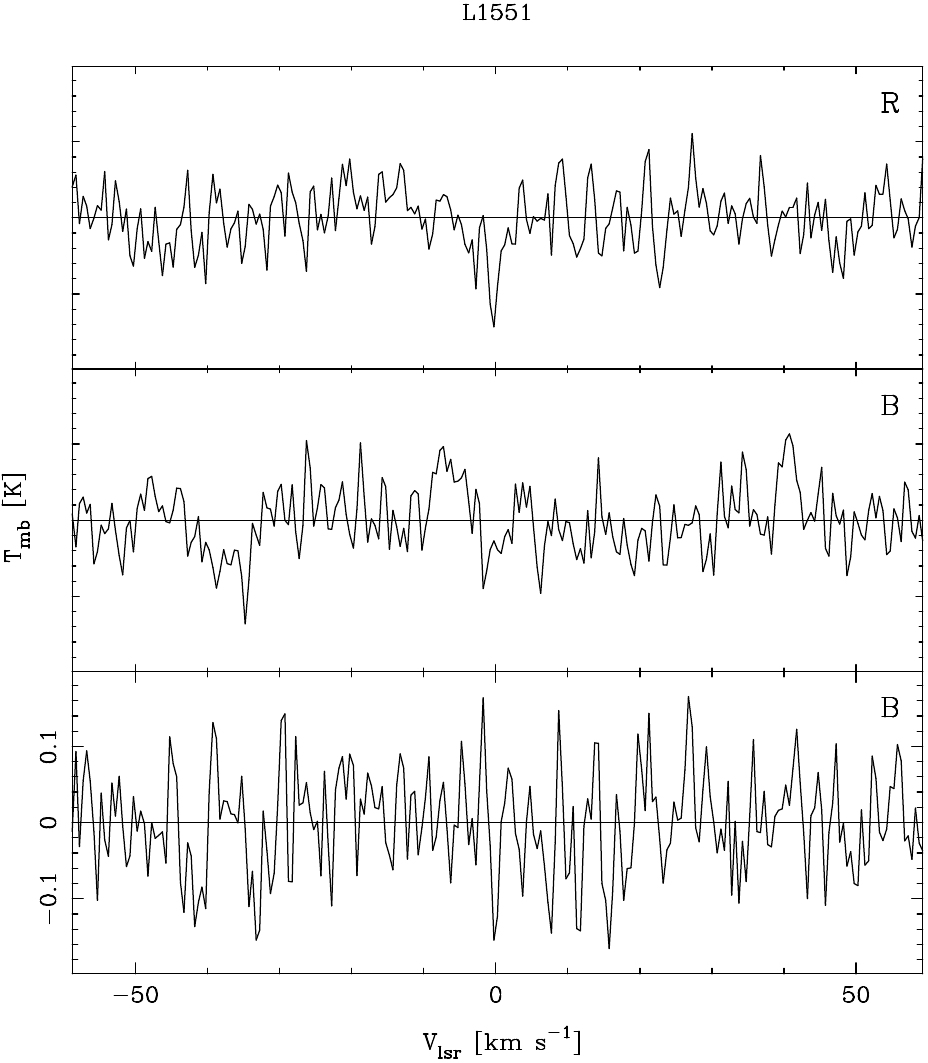}
   \\ \vspace{1cm}
   \includegraphics[width=7.5cm]{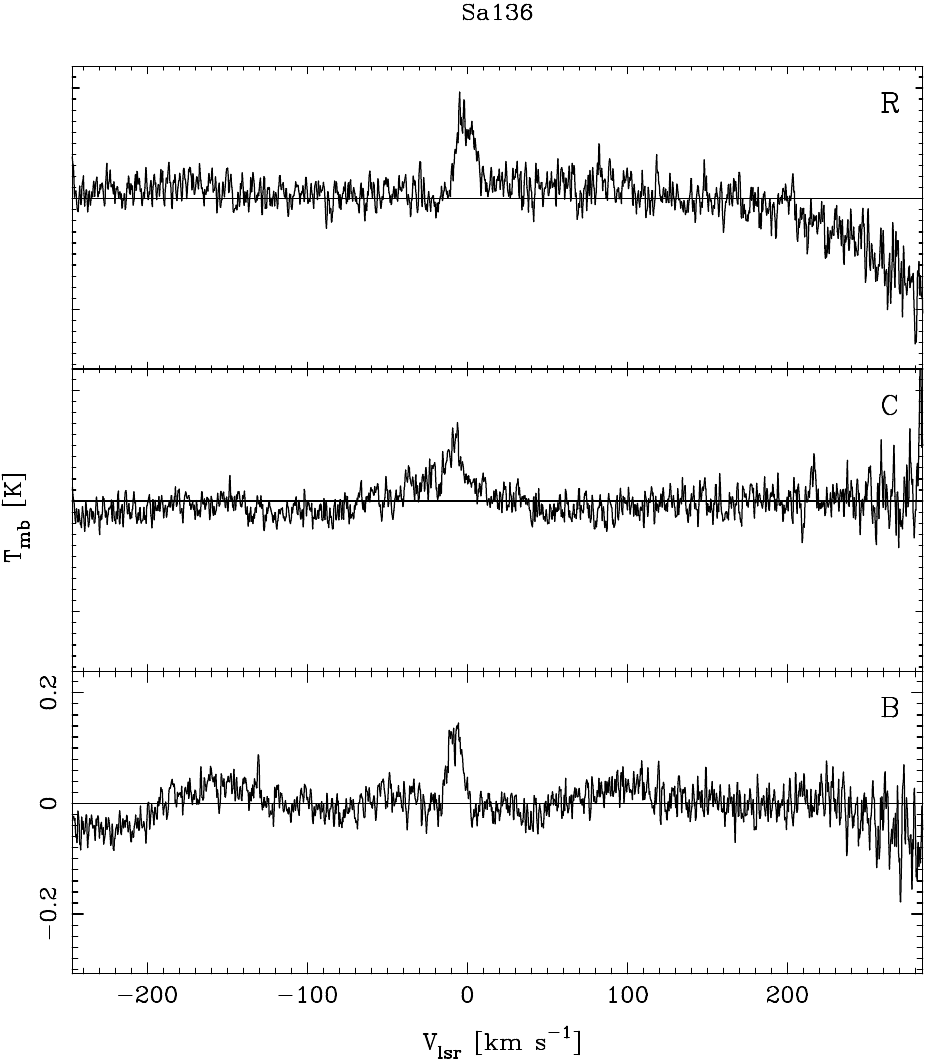} 
 \hspace{2cm}
 \includegraphics[width=7.5cm]{figureSa136.pdf}
 \\ \vspace{1cm}
   \includegraphics[width=7.5cm]{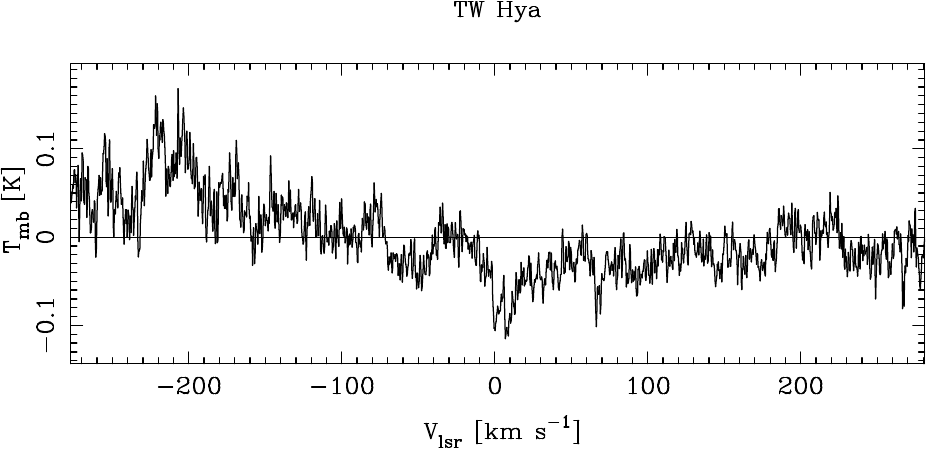} 
 \hspace{2cm}
 \includegraphics[width=7.5cm]{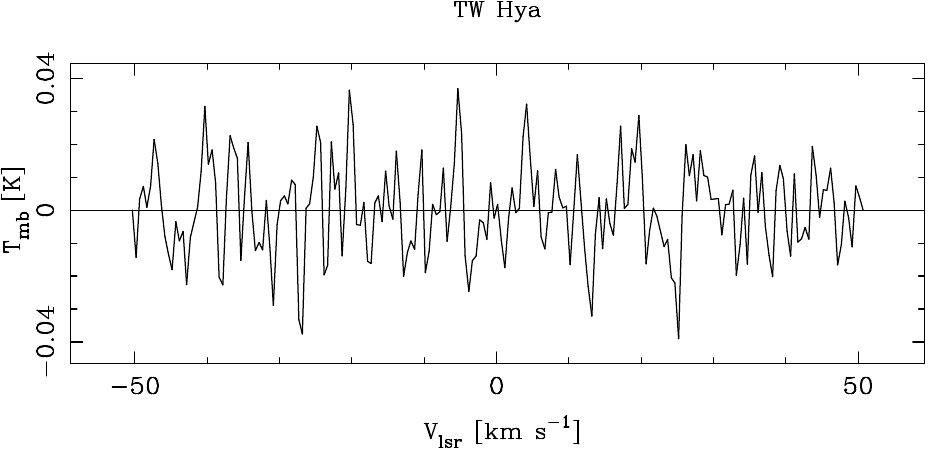}
     \caption{The same as Figure~\ref{fig:waterspectra} but for L1551, Sa136 and TW Hya.}
     \label{fig:waterspectra2}
\end{flushleft}
\end{figure*}

\begin{figure*}
\begin{flushleft}
 \includegraphics[width=7.5cm]{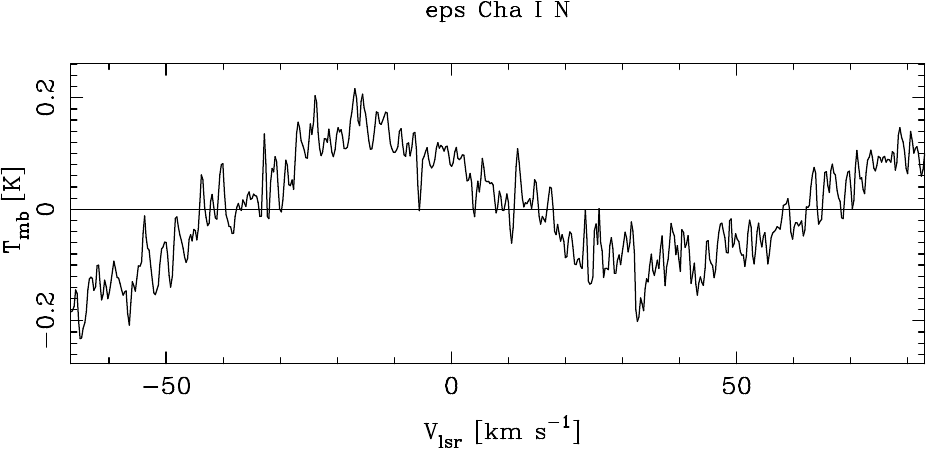} 
 \hspace{2cm}
 \includegraphics[width=7.5cm]{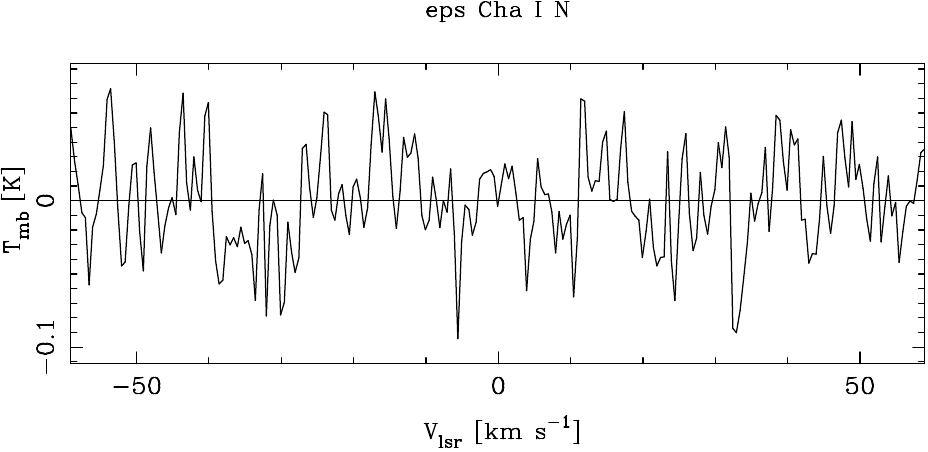}
   \\ \vspace{1cm}
   \includegraphics[width=7.5cm]{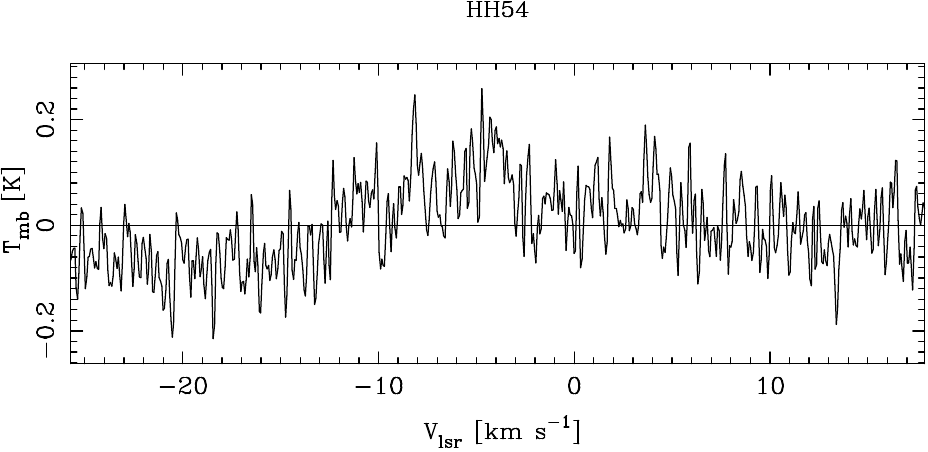} 
 \hspace{2cm}
 \includegraphics[width=7.5cm]{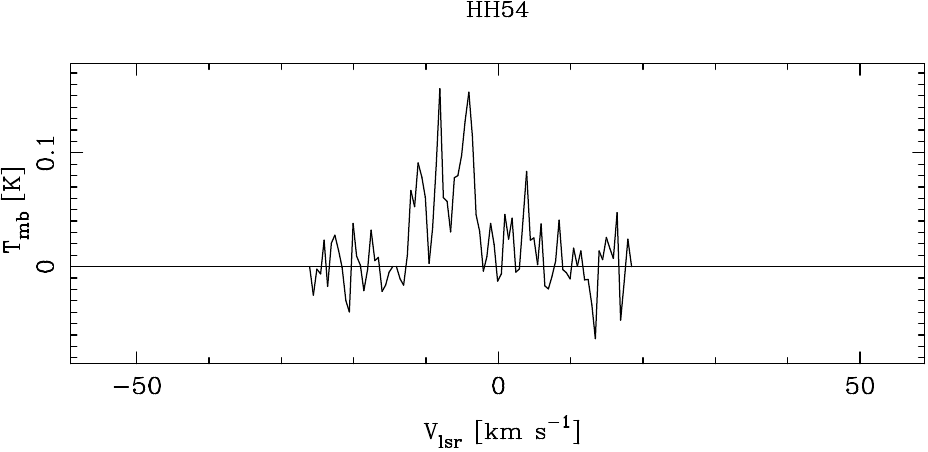}
 \\ \vspace{1cm}
   \includegraphics[width=7.5cm]{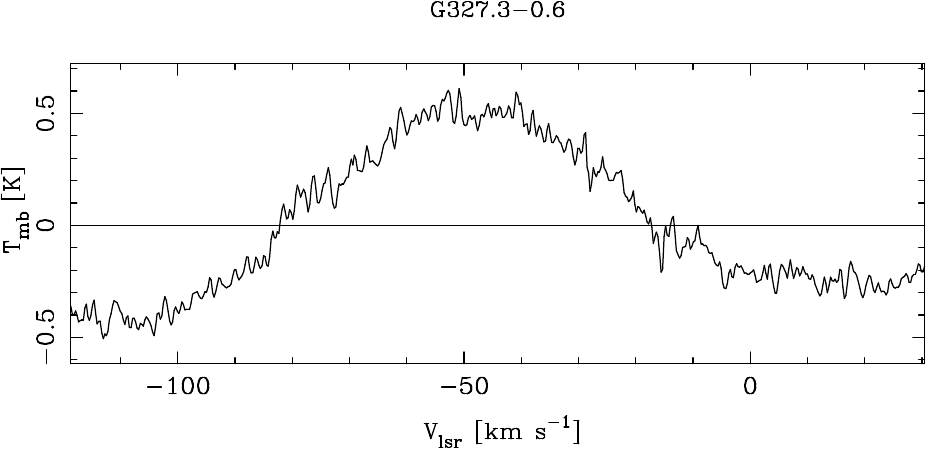} 
 \hspace{2cm}
 \includegraphics[width=7.5cm]{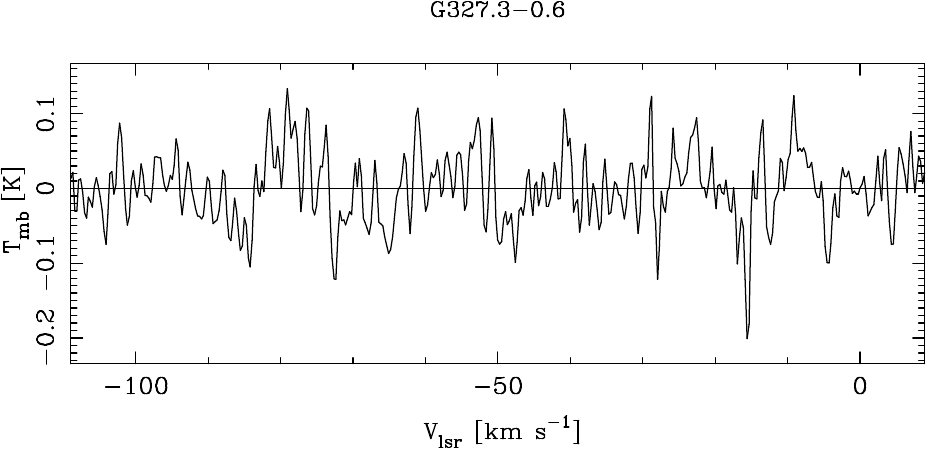}
\\ \vspace{1cm}
\includegraphics[width=7.5cm]{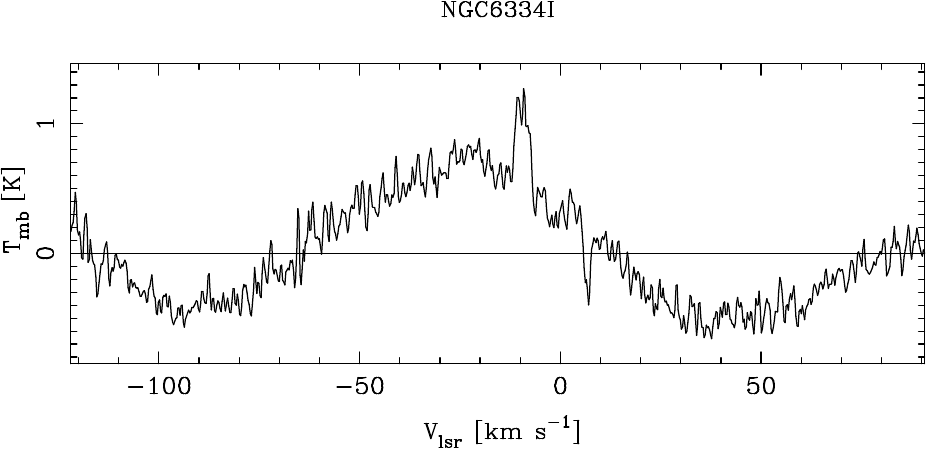} 
 \hspace{2cm}
 \includegraphics[width=7.5cm]{figureNGC6334I.pdf}
   \\ \vspace{1cm}
   \includegraphics[width=7.5cm]{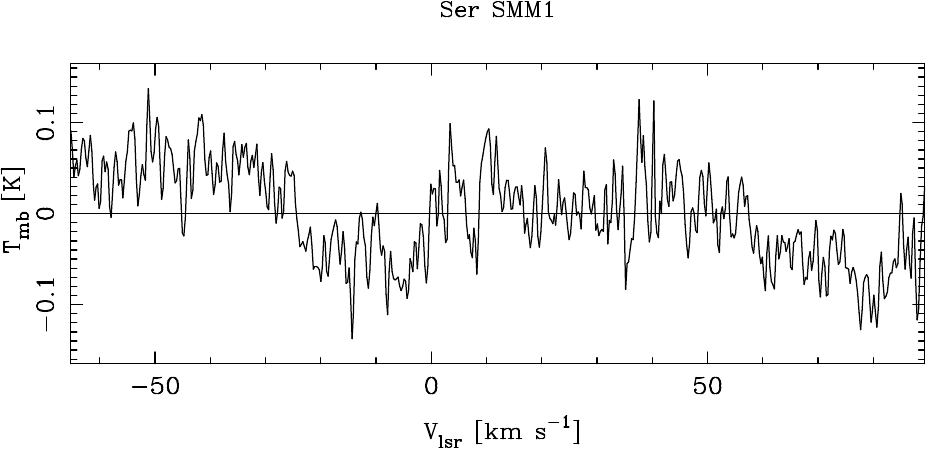} 
 \hspace{2cm}
 \includegraphics[width=7.5cm]{figureSer1VLA.pdf}
 \caption{The same as Figure~\ref{fig:waterspectra} but for eps Cha\,{\sc I\,}N, HH54 B, G327.3-0.6, NGC6334\,{\sc I} and Ser SMM1.}
     \label{fig:waterspectra3}
\end{flushleft}
\end{figure*}

\begin{figure*}
\begin{flushleft}
 \includegraphics[width=7.5cm]{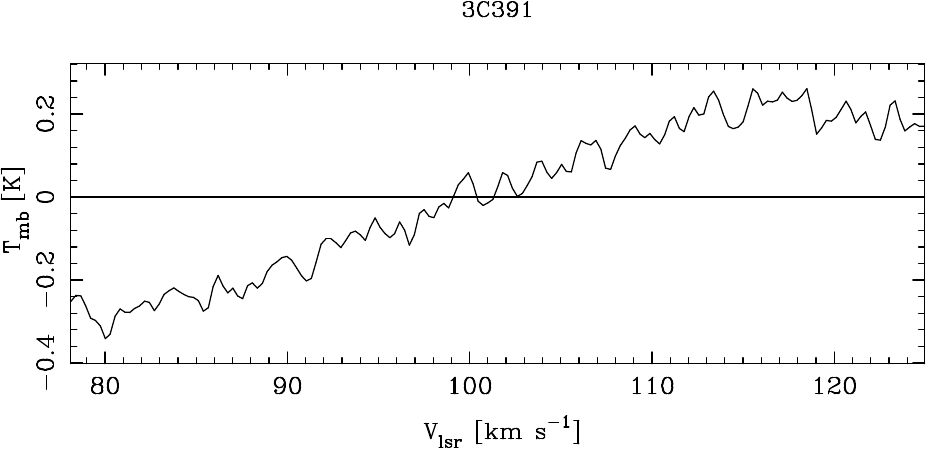} 
 \hspace{2cm}
 \includegraphics[width=7.5cm]{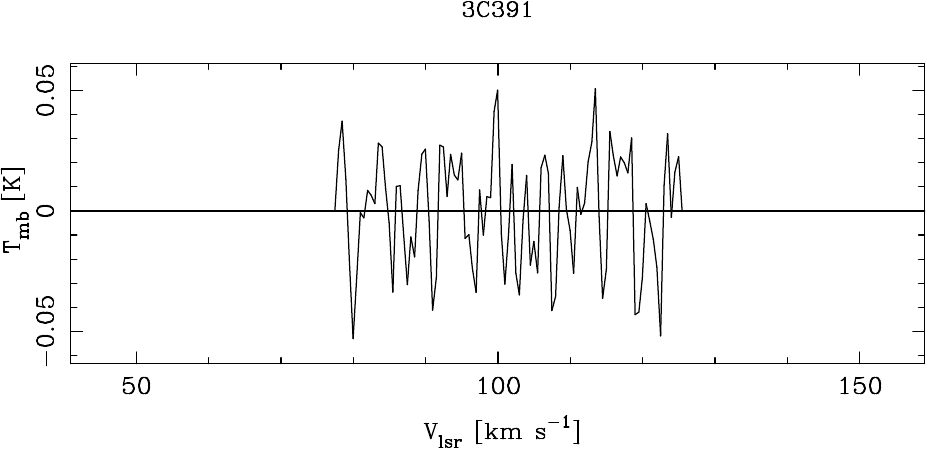}
   \\ \vspace{1cm}
   \includegraphics[width=7.5cm]{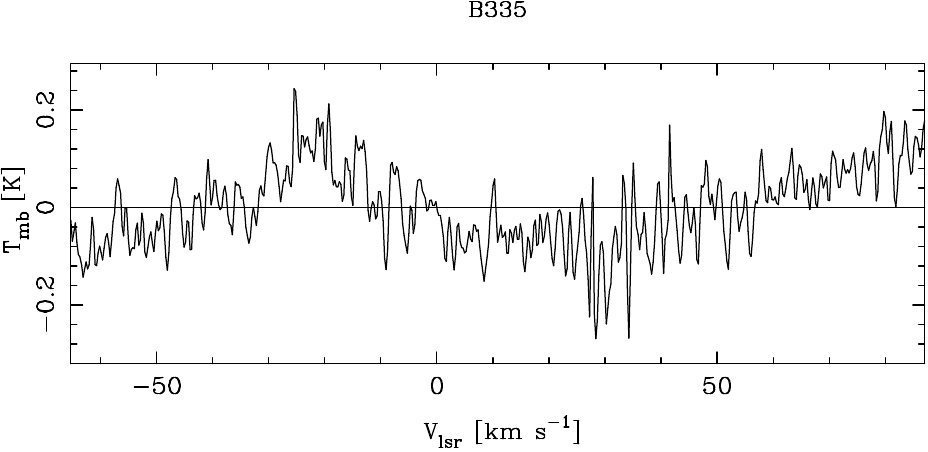} 
 \hspace{2cm}
 \includegraphics[width=7.5cm]{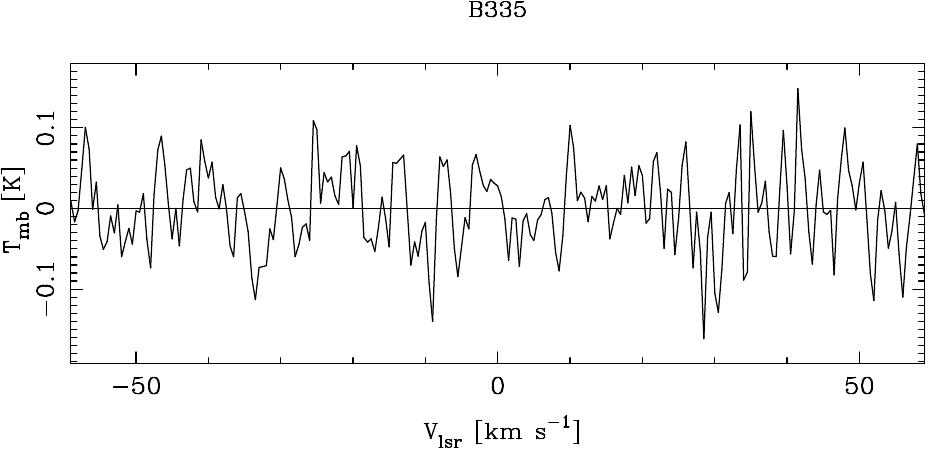}
 \\ \vspace{1cm}
   \includegraphics[width=7.5cm]{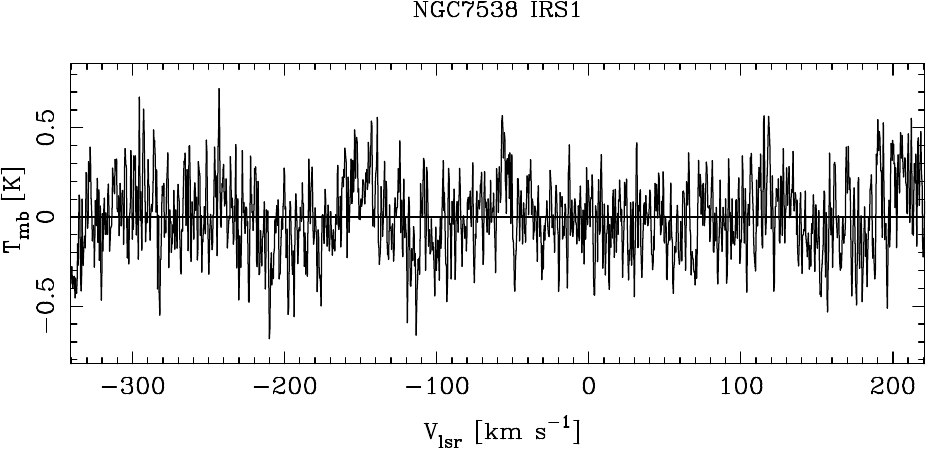} 
 \hspace{2cm}
 \includegraphics[width=7.5cm]{figureNGC7538.pdf}
\\ \vspace{1cm}
\includegraphics[width=7.5cm]{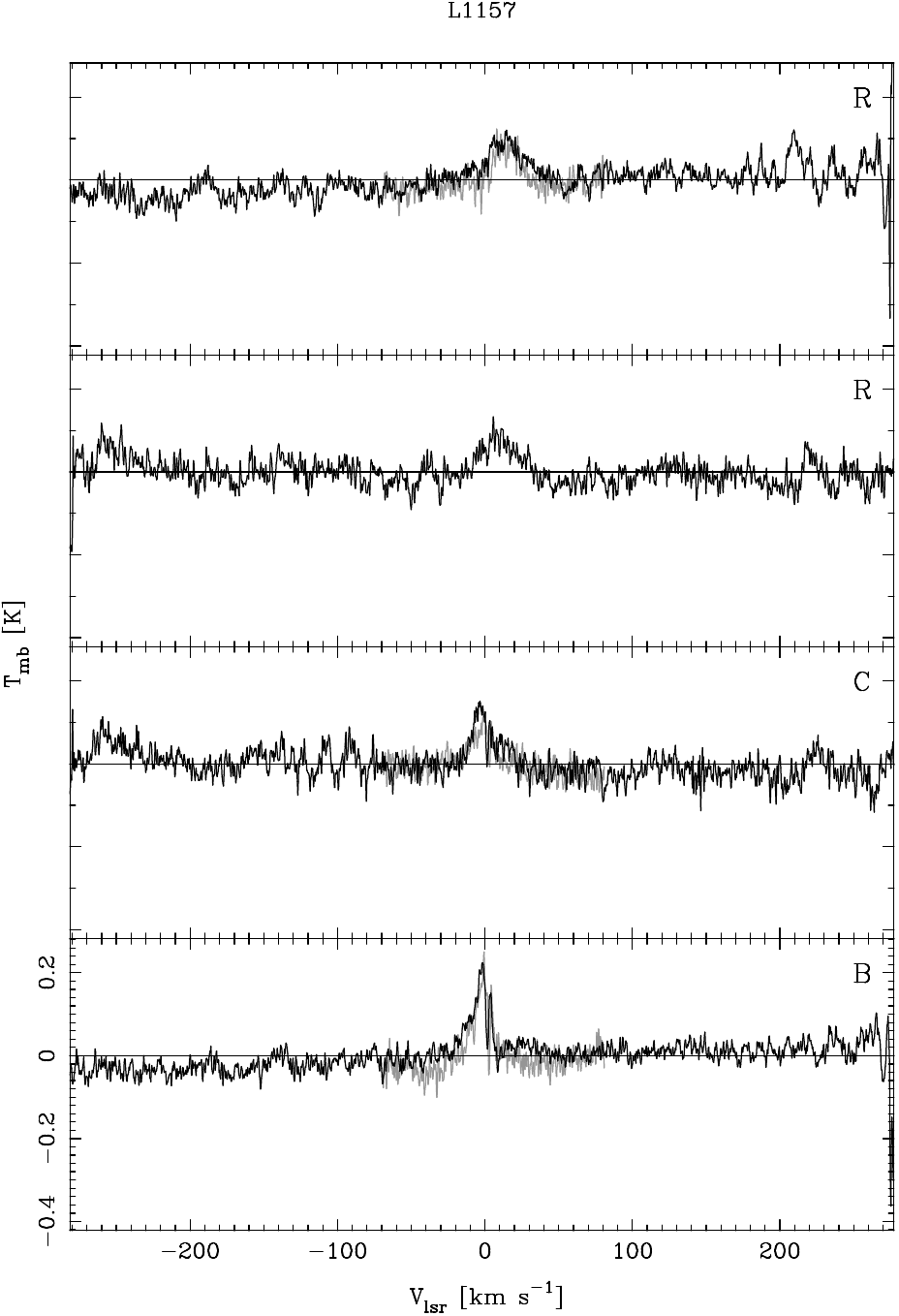} 
 \hspace{2cm}
 \includegraphics[width=7.5cm]{figureL1157.pdf}
 \caption{The same as Figure~\ref{fig:waterspectra} but for 3C391 BML,
   B335, NGC7538 IRS1 and L1157. For L1157, the AC2 data are plotted
   in gray and the AOS data in black.}
     \label{fig:waterspectra4}
\end{flushleft}
\end{figure*}

\end{appendix}

\end{document}